\documentclass[preprint,onecolumn,nofootinbib]{revtex4}
\usepackage[colorlinks=true,linkcolor=blue,urlcolor=blue,filecolor=black,citecolor=red,pdfstartview=FitV,pdftitle={},pdfsubject={},pdfkeywords={},pdfpagemode=None,bookmarksopen=true]{hyperref}
\usepackage{graphicx}
\usepackage{amsmath}
\usepackage{amsfonts}
\usepackage{amssymb,ulem}
\usepackage{color,xcolor}%
\usepackage{CJK}
\usepackage{subfigure}
\usepackage{amsthm,amsmath,amssymb}
\usepackage{mathrsfs}

\usepackage{dcolumn}
\usepackage{float}
\usepackage{multirow}
\begin{document}
\title{Transport properties in the Horndeski holographic two-currents model}
\author{Dan Zhang $^{1}$}
\thanks{danzhanglnk@163.com}
\author{Guoyang Fu$^{2}$}
\thanks{FuguoyangEDU@163.com}
\author{Xi-Jing Wang$^{2}$}
\thanks{xijingwang@yzu.edu.cn}
\author{Qiyuan Pan$^{1}$}
\thanks{panqiyuan@hunnu.edu.cn}
\author{Jian-Pin Wu$^{2}$}
\thanks{jianpinwu@yzu.edu.cn, corresponding author}
\affiliation{$^1$~ Key Laboratory of Low Dimensional Quantum Structures and Quantum Control of Ministry of Education, Synergetic Innovation Center for Quantum Effects and Applications, and Department of Physics, Hunan Normal University, Changsha, Hunan 410081, China}
\affiliation{$^2$~Center for Gravitation and Cosmology, College of Physical Science and Technology, Yangzhou University, Yangzhou 225009, China}
	
\begin{abstract}

The transport features of the holographic two-currents model are investigated in the Horndeski gravity framework.
This system displays metallic or insulating characteristics depending on whether the Horndeski coupling parameter $\gamma$ is negative or positive, but is unaffected by other system parameters such as the strength of the momentum dissipation $\hat{k}$, the doping $\chi$ and the coupling between two gauge fields $\theta$.
Secondly, we demonstrate that the thermal conductivities are affected not only by the inherent properties of the black hole, but also by the model parameters. Furthermore, we are particularly interested in the Lorentz ratios' properties. As expected, the Wiedemann-Franz (WF) law is violated, as it is in the majority of holographic systems. Particularly intriguing is the fact that several Lorentz ratio bounds reported in the typical axions model still remain true in our current theories. We would like to highlight out, however, that the lower bound for $\hat{\bar{L}}_A$ is affected by
the system parameters $\chi$, $\theta$ and $\gamma$, which differs from the case of the typical axions model.

\end{abstract}
	
\maketitle
\tableofcontents
	
\section{Introduction}

Transport is one of the most essential properties of strongly correlated systems in which the typical perturbative approach based on a single particle approximation loses power. A powerful tool for tackling these issues is the AdS/CFT correspondence, which connects a weakly coupled gravitational theory to a strongly coupled quantum field theory without gravity in the large N limit \cite{Maldacena:1997re,Gubser:1998bc,Witten:1998qj,Aharony:1999ti,Hartnoll:2016apf,Natsuume:2014sfa,Baggioli:2019rrs,Zaanen:2015oix}. The introduction of the momentum dissipation mechanism, which eliminates the $\delta$-function arising in the alternating current (AC) electric conductivity at zero frequency in the holographic system, represents a big step toward simulating more realistic systems.
The inclusion of momentum dissipation allows us to address strange metal behaviors \cite{Hartnoll:2009ns,Davison:2013txa,Donos:2013eha,Andrade:2013gsa,Baggioli:2021xuv,Horowitz:2012ky}, such as the linear-T resistivity and quadratic-T inverse Hall angle \cite{Blake:2014yla,Zhou:2015dha}, the universal properties of coherent and incoherent metals \cite{Davison:2015bea,Zhou:2015qui,Kim:2014bza}, the implementation of holographic metal-insulator transition (MIT) as well as the associated mechanism \cite{Donos:2012js,Donos:2014oha,Donos:2013eha,Donos:2014uba,Ling:2014saa,Baggioli:2014roa,Kiritsis:2015oxa,Ling:2015epa,Ling:2015exa,Ling:2016dck,Mefford:2014gia,Baggioli:2016oju,Andrade:2017ghg,Bi:2021maw}, etc.

The holographic framework provides various ways to introduce momentum dissipation, such as incorporating a spatially-dependent source into the dual boundary theory \cite{Horowitz:2012ky,Horowitz:2012gs,Ling:2013aya,Ling:2013nxa,Donos:2014yya}, utilizing Q-lattices \cite{Donos:2013eha,Donos:2014uba,Ling:2015epa,Ling:2015exa} or helical lattices \cite{Donos:2012js}, and introducing spatially linear dependent axion fields \cite{Andrade:2013gsa,Taylor:2014tka,Kim:2014bza,Cheng:2014tya,Ge:2014aza,Andrade:2016tbr,Kuang:2016edj,Tanhayi:2016uui,Kuang:2017cgt,Cisterna:2017jmv,Cisterna:2017qrb,Cisterna:2018hzf,Baggioli:2021ejg,Baggioli:2021xuv}. Among these mechanisms, the holographic axion model stands out for its versatility, simplicity, and efficiency. When using effective holographic low energy theories as a guide, it is natural and intriguing to investigate the impact of higher-derivative terms of axion fields \cite{Gouteraux:2016wxj,Baggioli:2016oqk,Baggioli:2016pia,Li:2018vrz,Huh:2021ppg}. Notably, within this holographic effective framework, it is possible to achieve spontaneous symmetry breaking and pseudo-spontaneous breaking \cite{Li:2018vrz,Ammon:2019wci,Baggioli:2019rrs,Alberte:2017oqx,Baggioli:2020ljz,Baggioli:2021xuv,Zhong:2022mok}.

The motivation for exploring these models stems from the challenge of reproducing and comprehending the exotic transport properties of strange metals. 
Furthermore, in numerous condensed matter systems, the fixed point is typically characterized by a non-trivial Lifshitz scaling exponent, which have been geometrically realized, as seen in \cite{Hertz:1976zz,Kachru:2008yh}. The whole gravity dual dictionary with Lifshitz asymptotics has also been formalized in \cite{Chemissany:2014xsa}. Given these considerations, it is highly recommended to investigate more general and intricate gravitational models. Horndeski gravity, as described by \cite{Horndeski:1974wa,Kobayashi:2019hrl}, is the most comprehensive scalar-tensor theory in four dimensions. Despite featuring higher-derivative terms beyond second-order in its Lagrangian, the theory's equations of motion remain second-order, ensuring that Horndeski gravity is ghost-free. This property is similar to Lovelock gravity \cite{Lovelock:1971yv}. In this regard, Horndeski theory is a particularly appealing framework due to its status as a natural extension of the holographic axion model \cite{Baggioli:2021ejg}.

The holographic applications of Horndeski gravity have been extensively studied in various works such as \cite{Kuang:2016edj,Caceres:2017lbr,Liu:2017kml,Li:2018kqp,Liu:2018hzo,Filios:2018xvy,Feng:2018sqm,Li:2018rgn,Santos:2021orr,Baggioli:2021ejg}, with a particular focus on its transport properties, as investigated in \cite{Jiang:2017imk,Baggioli:2017ojd,Wang:2019jyw,Figueroa:2020tya}. It has been discovered that the Horndeski coupling drives a metal-semiconductor-like transition \cite{Jiang:2017imk,Wang:2019jyw}. Furthermore, the impact of this new deformation on the universality of some well-known bound proposals has been investigated. While most of these proposals have been found to hold true \cite{Baggioli:2017ojd,Figueroa:2020tya}, violations of the heat conductivity-to-temperature lower bound and the viscosity-to-entropy ratio have been observed \cite{Figueroa:2020tya}.

In this paper, we want to study the two-currents model in Horndeski gravity framework using holographic duality technics. In condensed matter physics, the two-currents model has been employed to investigate various phenomena, such as the impact of electron-hole imbalances \cite{PhysRevB.79.085415} and spin population imbalance in ferromagnets \cite{Fert:2008zz,_uti__2004}. We would like to emphasize that the roots of spintronics can be traced back to Mott's two-currents model, which describes the electric and spin motive forces using two $U(1)$ gauge fields \cite{Mott1936The,Mott1936The}. Recently, two-currents models in holographic framework have also gained increasingly attention, see \cite{Kiritsis:2015hoa,Baggioli:2015dwa,Huang:2020iyw,Bigazzi:2011ak,Iqbal:2010eh,Rogatko:2017tae,Rogatko:2020vtz,Seo:2016vks,Zhang:2020znl,Rogatko:2019sjn,Erdmenger:2011hp,Dutta:2013osl,Correa:2019ivh,Musso:2013ija,Hafshejani:2018svs,Alsup:2012kr,Alsup:2012ap} and references therein. In these models, a pair of bulk $U(1)$ gauge fields $A$ and $B$ are connected to two independently conserved currents in the dual boundary field theory. The mismatch of the two independent chemical potentials or charge densities induces the unbalance of numbers. In holography, Dirac fluid, forming in the graphene near charge neutrality, has been constructed using two gauge fields in the gravity bulk \cite{Seo:2016vks} (also see \cite{Rogatko:2017tae,Rogatko:2020vtz,Rogatko:2019sjn} and references therein).
Particularly, the presence of a new current can significantly increase the heat transport relative to the charge transport, resulting in a violation of the Wiedemann-Franz (WF) law \cite{Seo:2016vks}. This suggests strong correlation of this system. Moreover, an unbalanced holographic superconductor has also been constructed as reported in \cite{Bigazzi:2011ak}. Specifically, the authors in \cite{Alsup:2012kr,Alsup:2012ap} have demonstrated that turning on either an interaction between the Einstein tensor and scalar field or a magnetic interaction between the second $U(1)$ field and scalar field, an inhomogeneous solution exhibits a higher critical temperature than the homogeneous case in the low-temperature limit. This leads to the emergence of Larkin-Ovchinnikov-Fulde-Ferrel (LOFF) states, which are characterized by a space modulated order parameter that corresponds to electron pairs with nonzero total momentum.

The paper is organized as follows. In Sec.\ref{sec-background}, we describe briefly the holographic two-currents model in the Horndeski gravity framework and work out the black hole solution. In Sec.\ref{sec-Transport}, we first derive the transport coeffcients, and then investigate the properties of the electric conductivity, spin-spin conductivity and thermal conductivity. Furthermore, we study the Lorentz ratios and discuss the WF law. Sec.\ref{conclusion} summarizes our findings and comments. In addition, we include two appendices that list the EOMs (Appendix \ref{appendix}) and provide a comprehensive derivation of the DC conductivity (Appendix \ref{appendix-B}).
	
\section{Holographic Background}\label{sec-background}

The gravitational background for this holographic model is taken in the form
\begin{eqnarray}\label{Action}
S= \int d^{4}x \sqrt{-g} \left[ \kappa\left(R-2\Lambda - \frac{1}{4}F^{2} - \frac{1}{4}Y^{2} - \frac{\theta}{2}F_{\mu\nu}Y^{\mu\nu}\right)-\frac{1}{2}(\lambda g ^{\mu\nu}-\gamma G^{\mu\nu})\sum_{I=x,y}\partial_{\mu}\phi^{I}\partial_{\nu}\phi^{I}\right]\,,\nonumber
\
\\
\end{eqnarray}
where $\Lambda=-3$ is the cosmological constant. A non-minimal coupling is added between the Einstein tensor $G^{\mu\nu}\equiv R^{\mu\nu}-\frac{1}{2}R g^{\mu\nu}$ and the axionic fields that produce momentum dissipation. This coupling is known as the Horndeski coupling, and its strength is denoted by the symbol $\gamma$. 
To avoid the ghost problem, $\gamma$ shall satisfies $-\infty<\gamma\leq 1/3$ \cite{Jiang:2017imk}. We introduce two U(1) gauge fields in this theory. The ordinary Maxwell field strength is represented by 
$F=dA$, whereas the second gauge field is indicated by $Y=dB$. We are also interested in the coupling term between the ordinary Maxwell field and the second gauge field, as in Refs. \cite{Rogatko:2017tae,Rogatko:2019sjn,Zhang:2020znl}. The coupling strength is denoted by the symbol $\theta$.
From now on, we will set the coupling constants to $\kappa=\lambda=1$ for convenience.

As mentioned in the introduction, the inclusion of a second $U(1)$ gauge field is motivated by accurately modeling carrier flow of strongly correlated systems. This is particularly relevant in systems such as the Dirac fluid in graphene or Mott's model. By introducing an interaction between the two $U(1)$ gauge fields through the coupling parameter $\theta$, an additional degree of freedom is created. The distinguishing and significant feature of the two-currents model is the tensor structure of the transport coefficients, which has general entries \cite{Bigazzi:2011ak,Rogatko:2017tae,Rogatko:2019sjn,Rogatko:2020vtz,Zhang:2020znl}.

According to the top-down perspective \cite{Acharya:2016fge}, the second $U(1)$ gauge field is refered to as the hidden sector. The interaction term $\theta$, also known as the kinetic mixing term, describes interaction of the ordinary Maxwell field (i.e., the first $U(1)$ gauge field) and the hidden $U(1)$ gauge field. This type of term was first introduced in \cite{Holdom:1985ag} to explain the existence and subsequent integration of heavy bi-fundamental fields charged under the $U(1)$ gauge groups. For more discussion on this topic, also please refer to \cite{Rogatko:2017tae, Rogatko:2019sjn, Rogatko:2020vtz}.

Applying the variational approach to the action \eqref{Action}, we can derive the EOMs (for the details, please see Appendix \ref{appendix}). Then, to solve these EOMs, we take the following ansatz:
\begin{eqnarray}\label{ansatz}
	&&
	ds^2= -h(r)dt^2+  \frac{dr^{2}}{f(r)}+r^{2}dx^idx^j\,, 
	\ \nonumber
	\\
	&&
	A=A_t(r)dt\,,\ \  B=B_t(r)dt\,, \ \ \phi^I=k x^I.
\end{eqnarray}
In the preceding ansatz, we assumed the axionic fields' spatial linear dependence, which adds momentum dissipation, such that the equation for the axionic fields \eqref{eom-axion} is satisfied automatically. So we simply need to solve the remaining EOMs \eqref{EE} and \eqref{ME} to find the functions, which are given by
\begin{eqnarray}\label{BG_solution}
&&
\label{BG_solutionv1}
h(r)=U(r)f(r)\,,\,\,\,\,\,\, U(r)=e^{\frac{\gamma k^{2}}{2r^{2}}}\,,
\
\\
&&
f(r)=\frac{e^{-\frac{k^{2}\gamma}{4r^{2}}}}{4kr\sqrt{\gamma}}
\Bigg(2k\sqrt{\gamma}\left(e^{\frac{k^{2}\gamma}{4r^{2}}}r\left(2r^{2}+k^{2}(\gamma-1)\right)
-e^{\frac{k^{2}\gamma}{4r_h^2}}r_h\left(2r_h^2+k^{2}(\gamma-1)\right)\right)
\ \nonumber
\\
&&
+\sqrt{\pi}\left(q_{A}^{2}+q_{B}^{2}+2\theta q_{A}q_{B}-k^{4}\gamma(\gamma-1)\right)\left(erfi\left(\frac{k\sqrt{\gamma}}{2r}\right)-erfi\left(\frac{k\sqrt{\gamma}}{2r_h}\right)\right)\Bigg)\,,
\label{BG_solutionv2}
\
\\
&&
\label{BG_solutionv3}
A_{t}(r)=\mu-q_{A}\frac{\sqrt{\pi} erfi(\frac{\sqrt{\gamma}k}{2r})}{\sqrt{\gamma}k}\,,
\ 
\\
&&
\label{BG_solutionv4}
B_{t}(r)=\delta\mu-q_{B}\frac{\sqrt{\pi} erfi(\frac{\sqrt{\gamma}k}{2r})}{\sqrt{\gamma}k}\,,
\end{eqnarray}
where $erfi(x)$ represents the imaginary error function and is expressed as $erfi(x)=\frac{2}{\sqrt{\pi}}\int_{0}^{x}e^{t^{2}}dt$. 
$r_{h}$ is the black hole's horizon determined by $f(r_h)=0$. When $q_{B}=\theta=0$, $f(r)$ recovers the result of the holographic Horndeski theory with one gauge field \cite{Jiang:2017imk,Baggioli:2017ojd}. Furthermore, when $\gamma=0$, this will return to the one of the typical holographic axions model \cite{Andrade:2013gsa}.  $\mu$, $\delta\mu$, $q_A$ and $q_B$ are the chemical potentials and the charge densities of the dual field theory associated with the gauge fields $A$ and $B$, respectively. The regularity of the gauge fields $A$ and $B$ at the horizon gives the following relations
\begin{eqnarray}
\mu&=&q_{A}\frac{\sqrt{\pi} erfi(\frac{\sqrt{\gamma}k}{2r_{h}})}{\sqrt{\gamma}k}\,, \nonumber
\label{Mu}
\\
\delta\mu&=&q_{B}\frac{\sqrt{\pi} erfi(\frac{\sqrt{\gamma}k}{2r_{h}})}{\sqrt{\gamma}k}\,.
\label{DMu}
\end{eqnarray}
The Hawking temperature of this black hole is
\begin{eqnarray}
T=-\frac{f'(r_h)}{4\pi}=\frac{e^{\frac{k^2 \gamma}{4}}(12-2k^2-q_A^2-q_B^2-2q_Aq_B\theta)}{16\pi}.
\end{eqnarray}
By the scaling symmetry, we can set $r_{h}=1$. In addition, we focus on the canonical ensemble by setting the chemical potential $\mu$ as the scaling unit\footnote{It is worth noting that in earlier publications on the holographic Horndeski theory \cite{Jiang:2017imk,Baggioli:2017ojd,Wang:2019jyw}, they usually set the charge density $q_A$ as the scaling unit.}.
After the parameters $\gamma$ and $\theta$ are fixed, the black hole solution is characterized by three dimensionless parameters $\hat{T}=T/\mu$, $\hat{k}=k/\mu$ and $\chi=\delta\mu/\mu$, the latter of which is used to simulate the doping \cite{Kiritsis:2015hoa,Baggioli:2015dwa,Huang:2020iyw} or represents the strength of the unbalance \cite{Bigazzi:2011ak}.

\section{Transport properties}\label{sec-Transport}

In this section, we shall calculate the DC thermoelectric transports of the dual field theory following the procedure proposed in \cite{Donos:2014uba} (for more details, see \cite{Blake:2014yla,Baggioli:2017ojd,Donos:2014cya,Blake:2013bqa,Baggioli:2016pia,Ling:2016dck}). 
Due to the rotation invariance on the $x-y$ plane, we only analyze the transport behaviors in $x$-direction. We turn on the constant electric fields $E_{A_{x}}$ and $E_{B_{x}}$, as well as the temperature gradient $\nabla T$, which induces the thermal gradient $\zeta\equiv-\nabla T/T$. 
They generate the corresponding electric currents $J_A^x$ and $J_B^x$, and the heat current $Q^x$. Following the terminology in \cite{Bigazzi:2011ak}, we also refer to $J_B^x$ as spin current.
By the generalized Ohm's law, we can calculate the corresponding transport coefficients:
\begin{eqnarray}\label{ DC definition}
	\left(
	\begin{array}{ccc}
		\sigma_A &\ \alpha T  & \ \eta\\
		\alpha T &\ \bar{\kappa}  T &\ \beta T\\
		\eta &\ \beta T  &\  \sigma_B\\
	\end{array}
	\right)=
	\left(
	\begin{array}{ccc}
		\frac{\partial J_A^x }{\partial E_{Ax}} &\frac{\partial J_A^x }{\partial \zeta} & \frac{\partial J_A^x }{\partial E_{Bx}} \\
		\frac{\partial Q^x }{\partial E_{Ax}} & \frac{\partial Q^x }{\partial \zeta} & \frac{\partial Q^x }{\partial E_{Bx}} \\
		\frac{\partial J_B^x }{\partial E_{Ax}} &\frac{\partial J_B^x }{\partial \zeta} & \frac{\partial J_B^x }{\partial E_{Bx}}\\
	\end{array}
	\right)\,.
	\label{dceq}
\end{eqnarray}
$\sigma_A$ and $\sigma_B$ are the electric conductivity and spin-spin conductivity associated to the gauge fields $A$ and $B$, respectively.
$\eta$ is called the spin conductivity, which measures the spin current $J_B^x$ generated by the electric field $E_{A_{x}}$ even without $E_{B_{x}}$, or vice versa.  $\bar{\kappa}$ is the thermal conductivity induced by the heat current associated to the temperature gradient.
$\alpha$ and $\beta$ are the thermo-electric and thermo-spin conductivities, respectively. In the absence of the temperature gradient $\nabla T$, they can also be caused by the heat current with the electric fields $E_{A_{x}}$ and $E_{B_{x}}$.

Following the strategy outlined in \cite{Donos:2014cya} (also see \cite{Baggioli:2017ojd}), we can work out the DC conductivities. Appendix \ref{appendix-B} has the complete derivation. For convenience, we've also copied them here:
\begin{eqnarray}\label{DC1}
	&&
	\label{sigmaA}
	\hat{\sigma}_{A}=1+\frac{\hat{k}^{2}\mu^{4}\gamma(1+\theta\chi)^{2}}{\hat{M_{h}^{2}}\pi erfi(\frac{\hat{k}\mu\sqrt{\gamma}}{2})^{2}} \,,	
	\\
	&&
	\label{sigmaB}
	\hat{\sigma}_{B}=1+\frac{\hat{k}^{2}\mu^{4}\gamma(\theta+\chi)^{2}}{\hat{M_{h}^{2}}\pi erfi(\frac{\hat{k}\mu\sqrt{\gamma}}{2})^{2}} \,, \\
	&&
	\label{alpha}
	\hat{\alpha}=\frac{4\hat{k}\mu^2\sqrt{\pi}\sqrt{\gamma}(1+\theta\chi)}{\hat{M_{h}^{2}} erfi(\frac{\hat{k}\mu\sqrt{\gamma}}{2})}\,, \\
	&&
	\label{beta}
	\hat{\beta}=\frac{4\hat{k}\mu^2\sqrt{\pi}\sqrt{\gamma}(\theta+\chi)}{\hat{M_{h}^{2}} erfi(\frac{\hat{k}\mu\sqrt{\gamma}}{2})}\,, \\
	&&
	\label{eta}
	\hat{\eta}=\theta+\frac{\hat{k}^{2}\mu^{4}\gamma(\theta+\chi)(1+\theta\chi)}{\hat{M_{h}^{2}}\pi erfi(\frac{\hat{k}\mu\sqrt{\gamma}}{2})^{2}}\,,\\
	&&
	\label{bar}
	\hat{\bar{\kappa}}=\frac{16\pi^{2}\hat{T}}{\hat{M_{h}^{2}}}\,. 
\end{eqnarray}	
Notice that the hat here, as well as throughout this paper, indicates dimensionless quantities. $\hat{M}_h$ is the effective graviton mass at the horizon (see Eq.\eqref{Mh}). In this research, we will primarily investigate the properties of the electric and spin-spin conductivities. The thermal conductivities are then briefly discussed. We are also interested in the Lorentz ratios, which are connected to the thermal and electric conductivities, and we investigate their characteristics in depth.

\subsection{Electric and spin-spin conductivities}
\begin{figure}[ht]
	\centering
	\subfigure{
		\includegraphics[width=6.6cm]{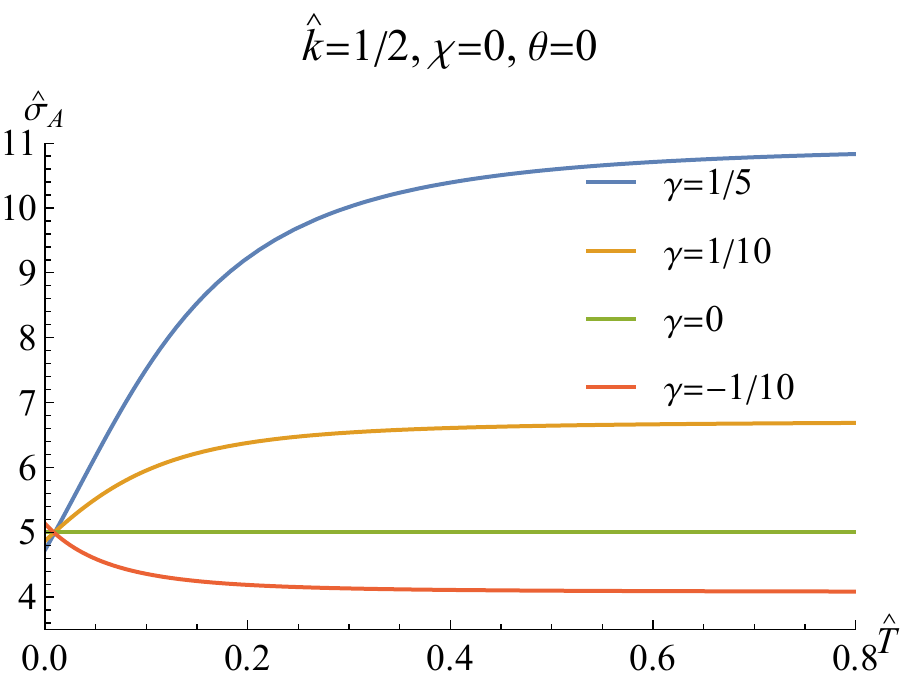}}
	\hspace{0.5in}
	\subfigure{
		\includegraphics[width=6.5cm]{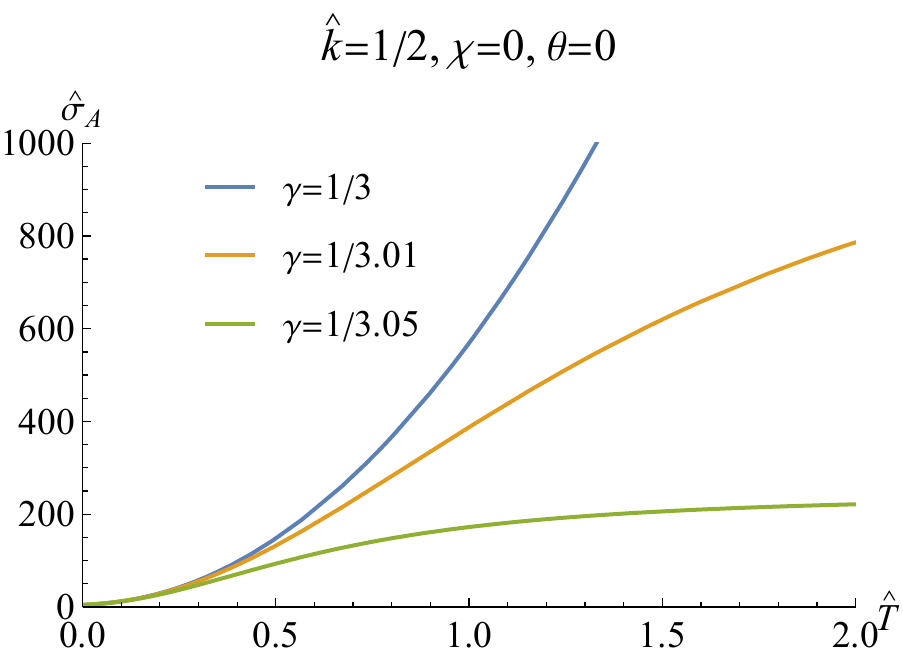}}
	\caption{\label{sigmaA-chiv0-thetav0} Electric conductivity $\hat{\sigma}_{A}$ as a function of the temperature $\hat{T}$, with specified $\hat{k}=1/2$ and varying different $\gamma$. Here, we've set $\chi=0, \theta=0$.}
\end{figure}
\begin{figure}[ht]
	\centering
	\subfigure{
		\includegraphics[width=6.5cm]{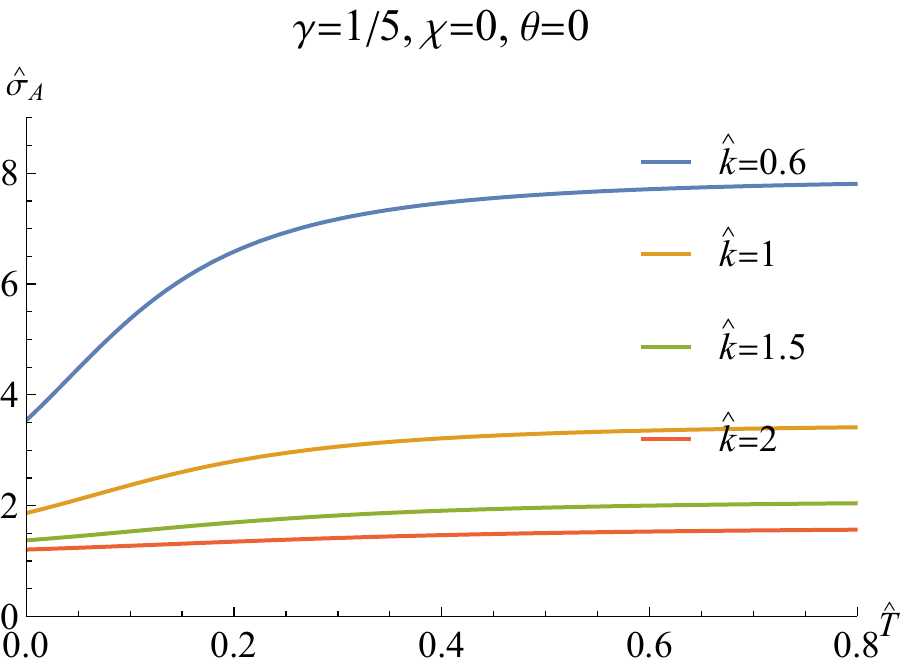}}
	\hspace{0.5in}
	\subfigure{
		\includegraphics[width=6.5cm]{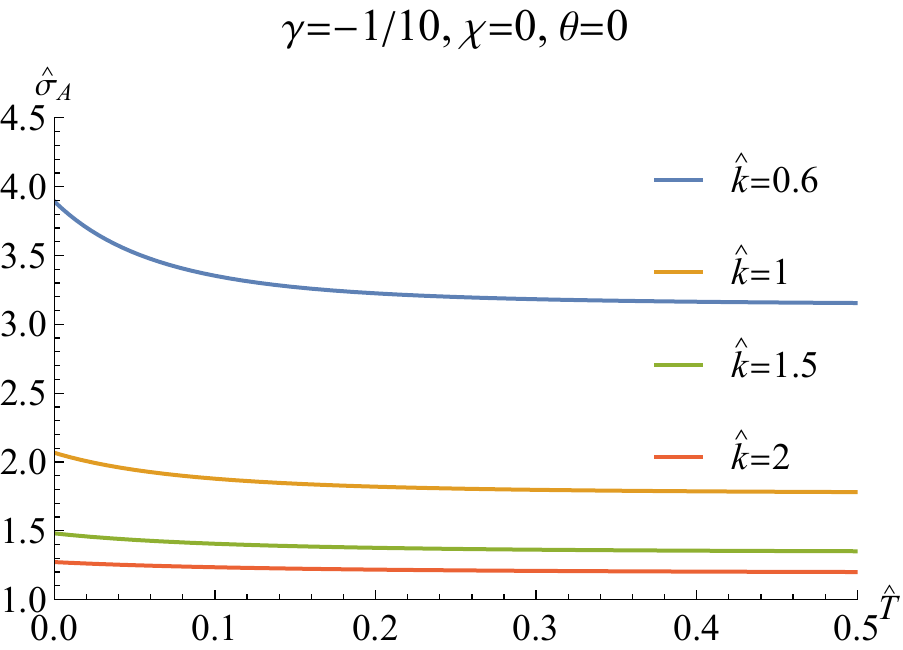}}
	\caption{\label{sigmaA-chiv0-thetav0-changingk} Electric conductivity $\hat{\sigma}_{A}$ as a function of the temperature $\hat{T}$ for varivd different $\hat{k}$ at $\gamma$ = $1/5$ and $-1/10$ . Here, we've set $\chi=0, \theta=0$.}
\end{figure}

We first investigate the properties of electric conductivity in the absence of doping, for which this theory reduces to the dual theory of the holographic Horndeski model with linear axionic fields explored in \cite{Jiang:2017imk,Baggioli:2017ojd,Wang:2019jyw}. 
To this purpose, we present the temperature behaviors of the DC electric conductivity $\hat{\sigma}_{A}$ for a given $\hat{k}$ and varied $\gamma$ in Fig.\ref{sigmaA-chiv0-thetav0}, and for a specified $\gamma$ and various $\hat{k}$ in Fig.\ref{sigmaA-chiv0-thetav0-changingk}.
Following is a summary of the key characteristics.
\begin{itemize}
	\item When $\gamma=0$, the system is further reduced to the typical holographic axions model \cite{Andrade:2013gsa}, and the DC electric conductivity is temperature independent (see the green line of left-plot in Fig.\ref{sigmaA-chiv0-thetav0}).
	\item When the value of $\gamma$ deviates from zero, the system displays metallic or insulating behaviors, depending on whether $\gamma$ is negative or positive.
	Particularly for $\gamma>0$, when temperature drops, the electric conductivity decreases, behaving like an insulator (Fig.\ref{sigmaA-chiv0-thetav0}). 
	For $\gamma<0$, the inverted behaviors arise, and the system is identified as a metal (left-plot in Fig.\ref{sigmaA-chiv0-thetav0}). We would like to emphasize that given $\gamma$, the dual system exhibits metallic or insulating characteristics that are independent of momentum dissipation strength (Fig.\ref{sigmaA-chiv0-thetav0-changingk}). This picture closely resembles the holographic axions model with non-linear Maxwell field \cite{Baggioli:2016oju,Wu:2018zdc} or gauge-axion coupling \cite{Baggioli:2016oqk,Gouteraux:2016wxj}. 
	\item We would like to mention that  MIT can be induced by the strength of the momentum dissipation in the holographic EMAW (Einstein-Maxwell-axion-Weyl) theory \cite{Ling:2016dck,Wu:2018pig}, where a higher-derivative term involving the coupling between the Weyl tensor and the Maxwell field is introduced. However, momentum dissipation in our current model merely suppresses electric conductivity and does not cause the MIT (Fig.\ref{sigmaA-chiv0-thetav0-changingk}).
	\item In the high temperature limit, electric conductivity approaches to infinity when $\gamma$ saturates the upper bound, i.e., $\gamma=1/3$. When $\gamma$ deviates from this bound, it tends to a constant in the hight temperautre limit (right-plot in Fig.\ref{sigmaA-chiv0-thetav0}).
\end{itemize}
\begin{figure}[ht]
	\centering
	\subfigure[]{
		\includegraphics[width=5.21cm]{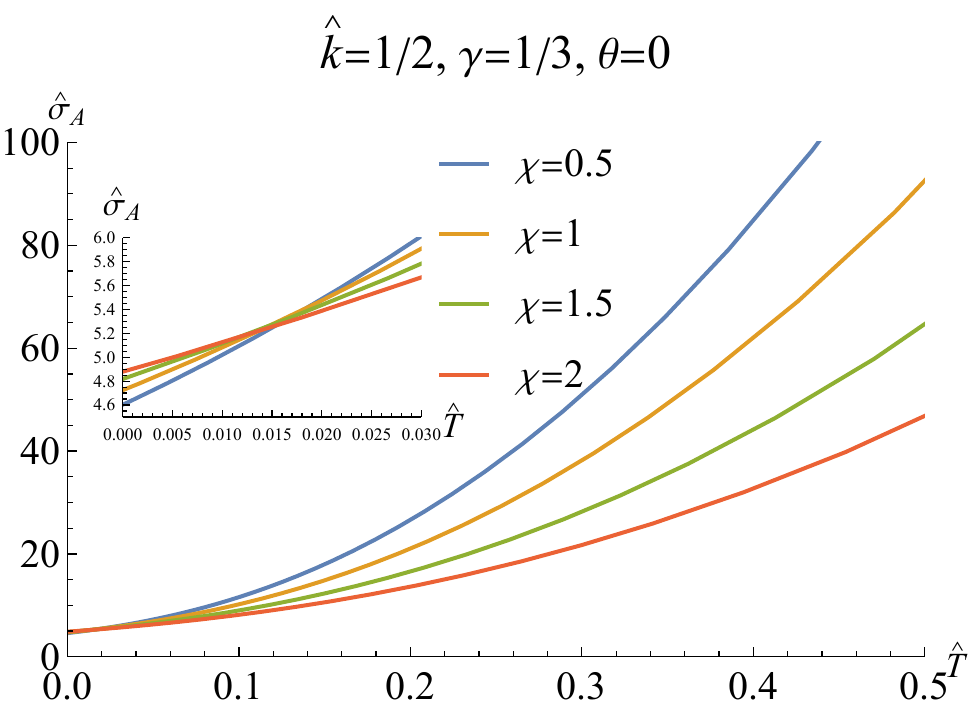}}
	\hspace{0in}
	\subfigure[]{
		\includegraphics[width=5.2cm]{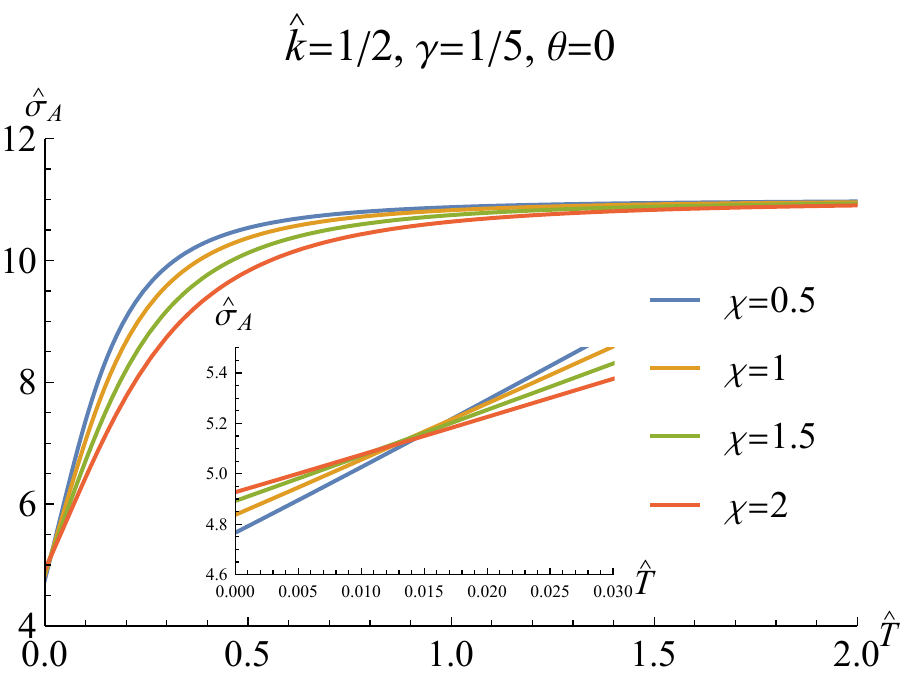}}
	\hspace{0in}
	\subfigure[]{
		\includegraphics[width=5.2cm]{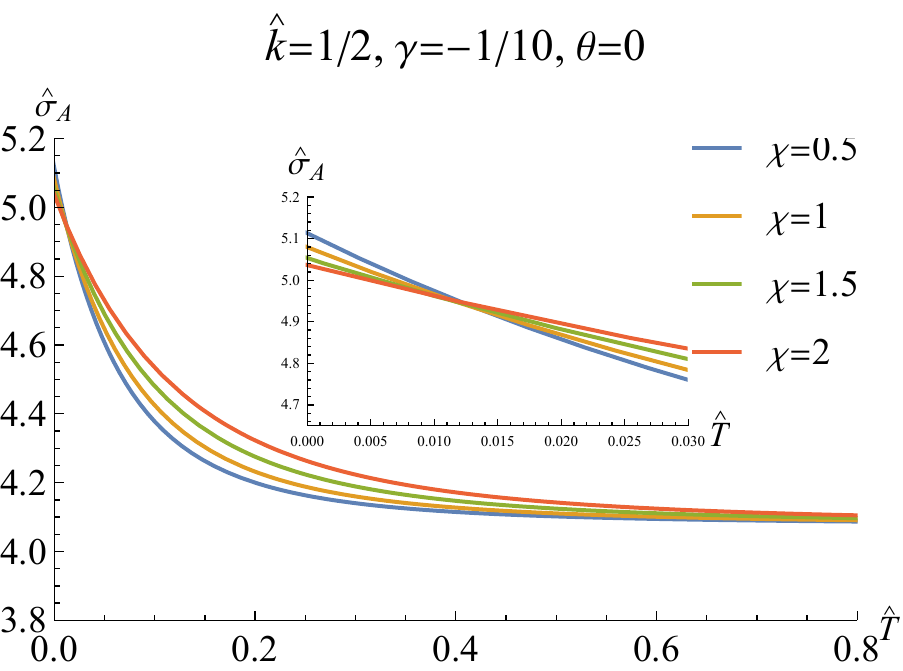}}
	\caption{\label{sigmaA-k1-thetav0-changingc} Electric conductivity $\hat{\sigma}_{A}$ as a function of the temperature $\hat{T}$ for various $\chi$ values at $\gamma$ = $1/3$, $1/5$ and $-1/10$. Here, we've set $\hat{k}=1/2, \theta=0$.}
\end{figure}
\begin{figure}[ht]
	\centering
	\subfigure[]{
		\includegraphics[width=5.2cm]{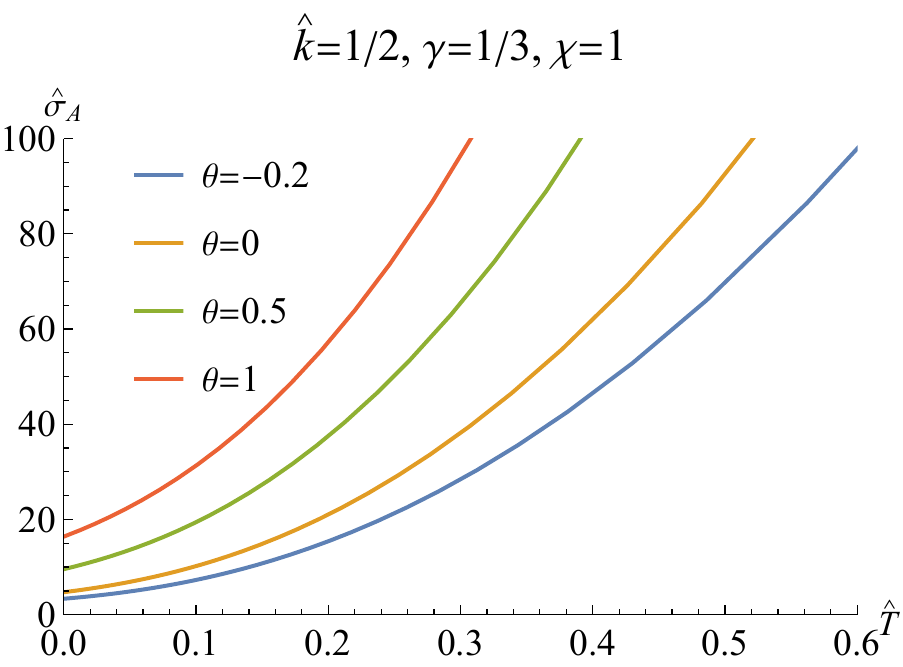}}
	\hspace{0in}
	\subfigure[]{
		\includegraphics[width=5.2cm]{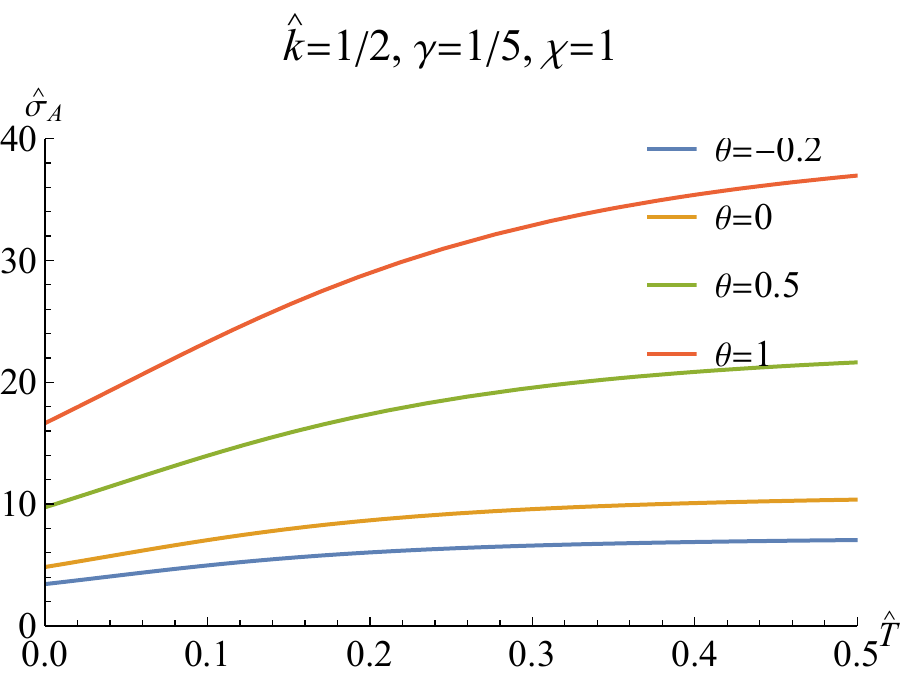}}
	\hspace{0in}
	\subfigure[]{
		\includegraphics[width=5.2cm]{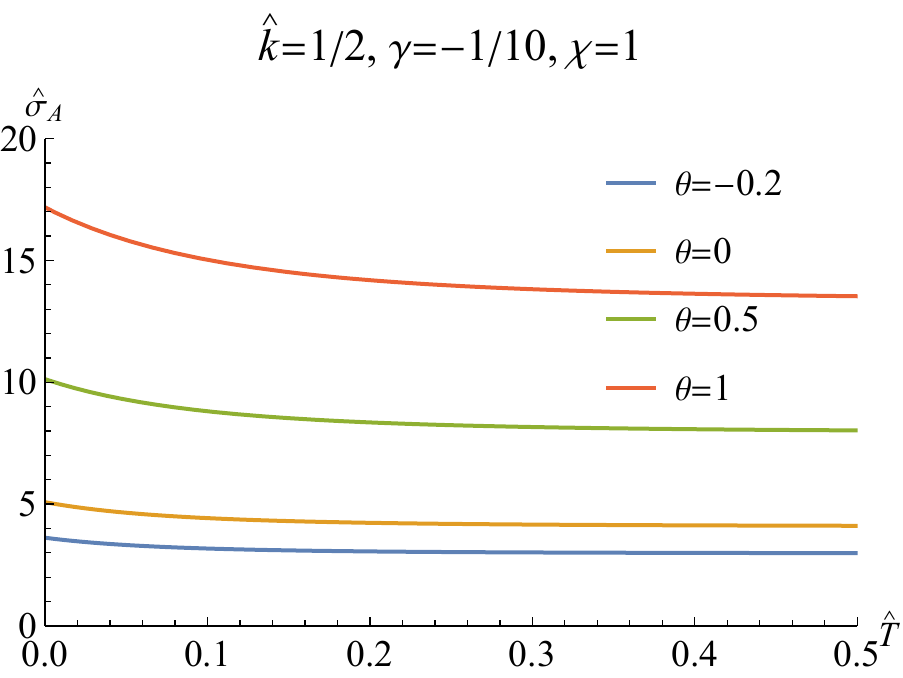}}
	\caption{\label{sigmaA-k1-thetav1-changingc} Electric conductivity $\hat{\sigma}_{A}$ as a function of the temperature $\hat{T}$ for various $\theta$ values at $\gamma$ = $1/3$, $1/5$ and $-1/10$. Here, we've set $\hat{k}=1/2, \chi=1$.}
\end{figure}

The effects of doping $\chi$ and coupling $\theta$ are then investigated. To that end, in Fig.\ref{sigmaA-k1-thetav0-changingc}, we present the temperature behaviors of electric conductivity with various $\chi$ for the selected $\gamma$ and $\hat{k}$, and various $\theta$ for the selected $\gamma$, $\chi$ and $\hat{k}$ in Fig.\ref{sigmaA-k1-thetav1-changingc}. The properties are summarized below.
\begin{itemize}
	\item Qualitatively, the electric conductivity increases or decreases with decreasing temperature, regardless of doping. However, doping has a distinct effect on electric conductivity at various temperatures. Electric conductivity reduces in the high temperature region as doping increases. An inverted behavior emerges in the low temperature region (see the inset in Fig.\ref{sigmaA-k1-thetav0-changingc}).
	\item Electric conductivity diminishes as $\theta$ decreases. The coupling $\theta$, on the other hand, cannot modify the temperature characteristics of electric conductivity.
\end{itemize}

\begin{figure}[ht]
	\centering
	\subfigure[]{
		\includegraphics[width=5.2cm]{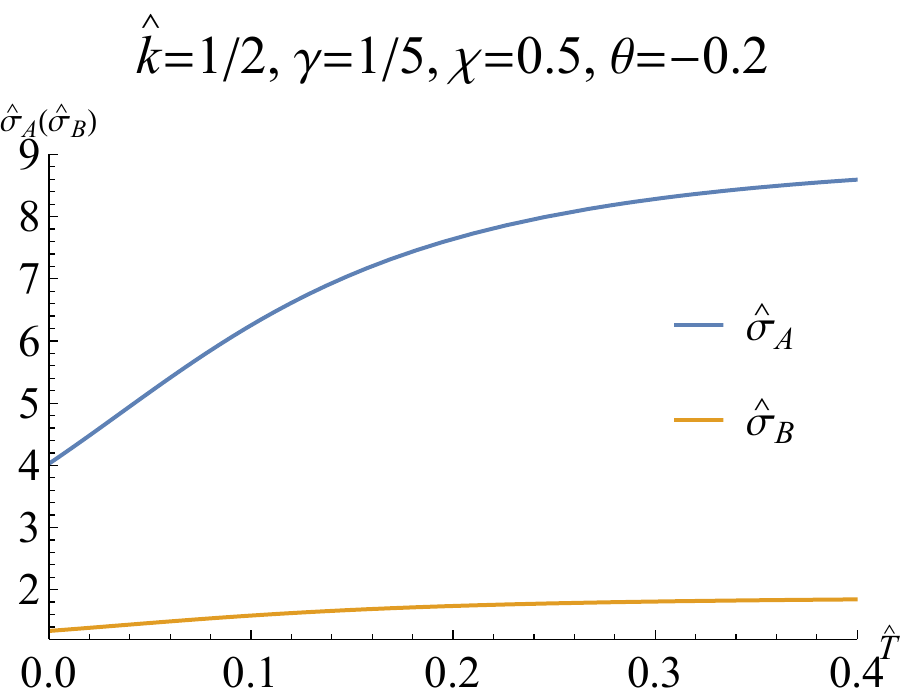}}
	\hspace{0in}
	\subfigure[]{
		\includegraphics[width=5.2cm]{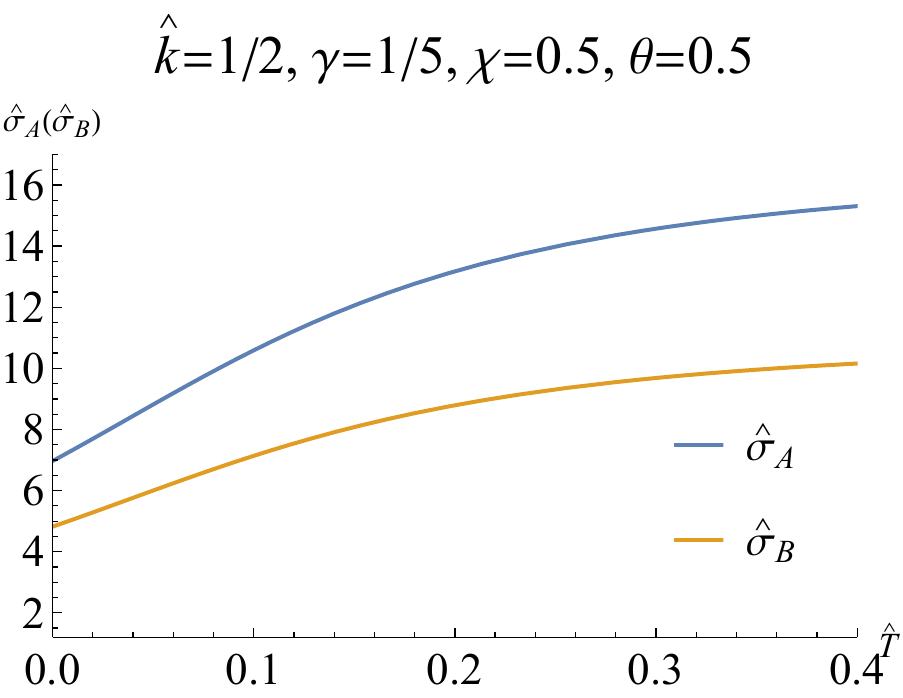}}
	\hspace{0in}
	\subfigure[]{
		\includegraphics[width=5.2cm]{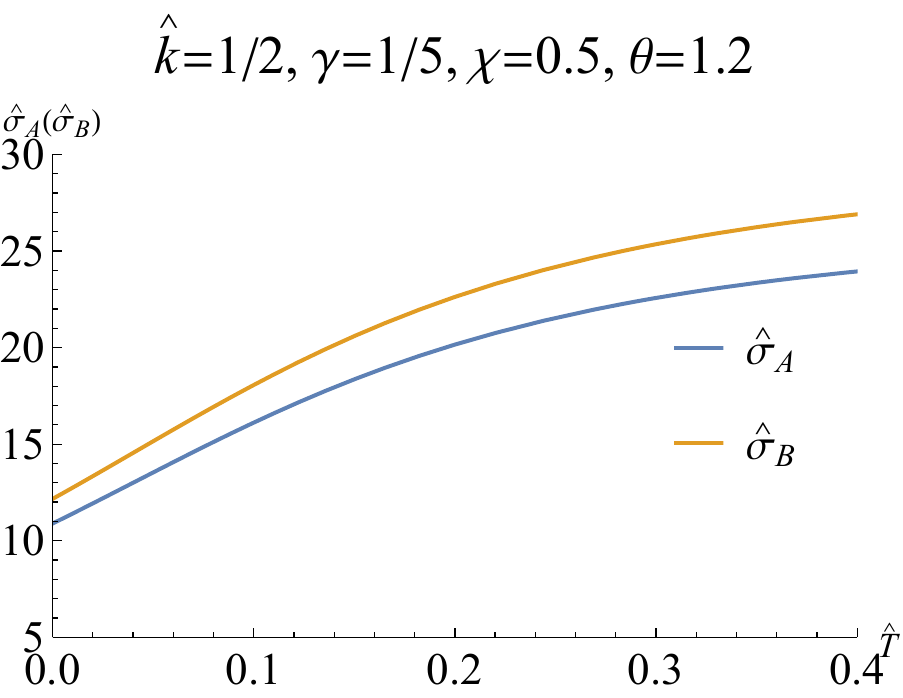}}
	\hspace{0in}
	\subfigure[]{
		\includegraphics[width=5.2cm]{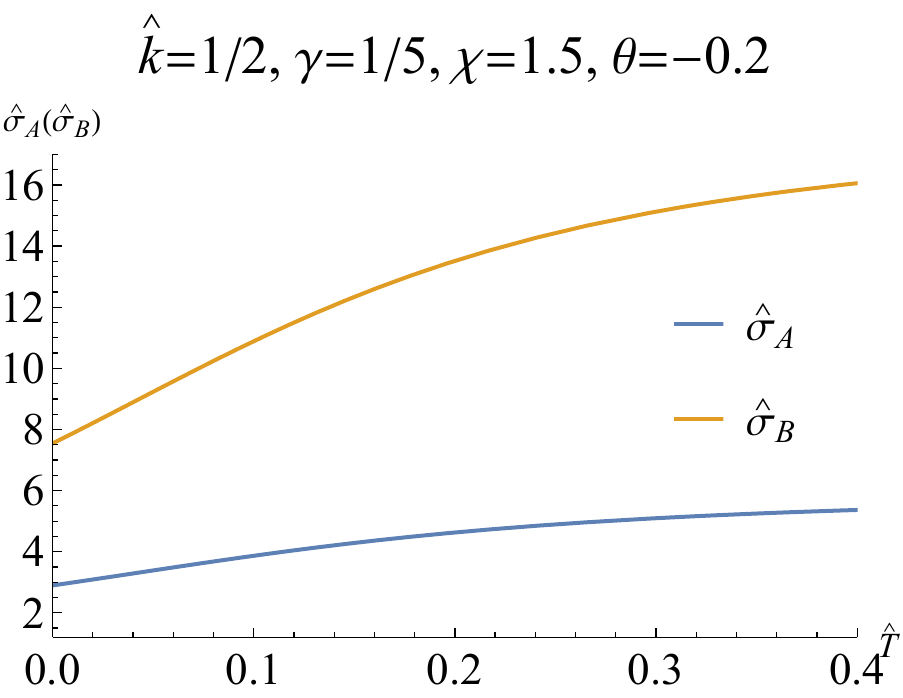}}
	\hspace{0in}
	\subfigure[]{
		\includegraphics[width=5.2cm]{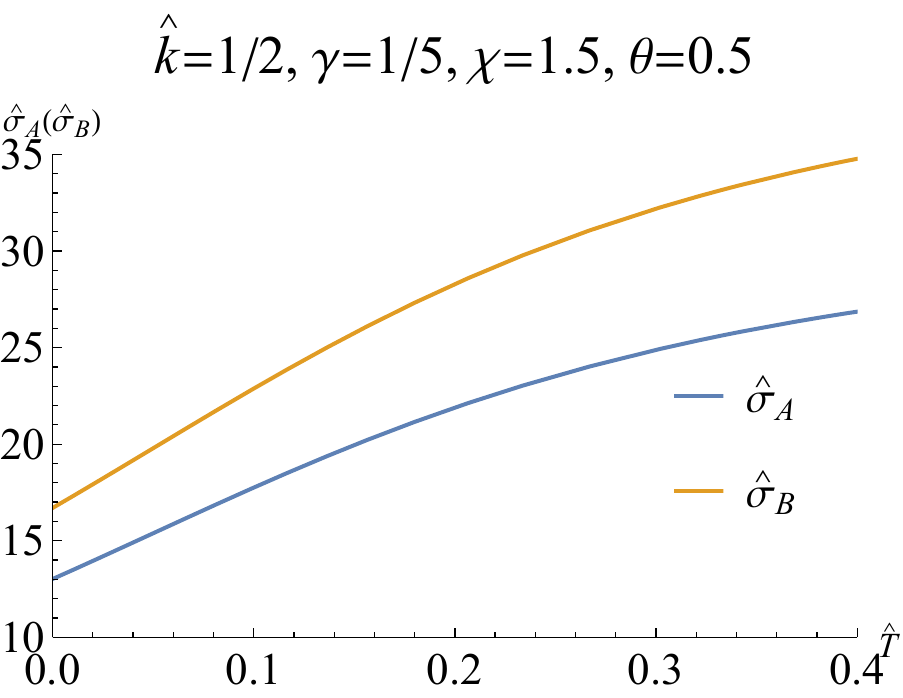}}
	\hspace{0in}
	\subfigure[]{
		\includegraphics[width=5.2cm]{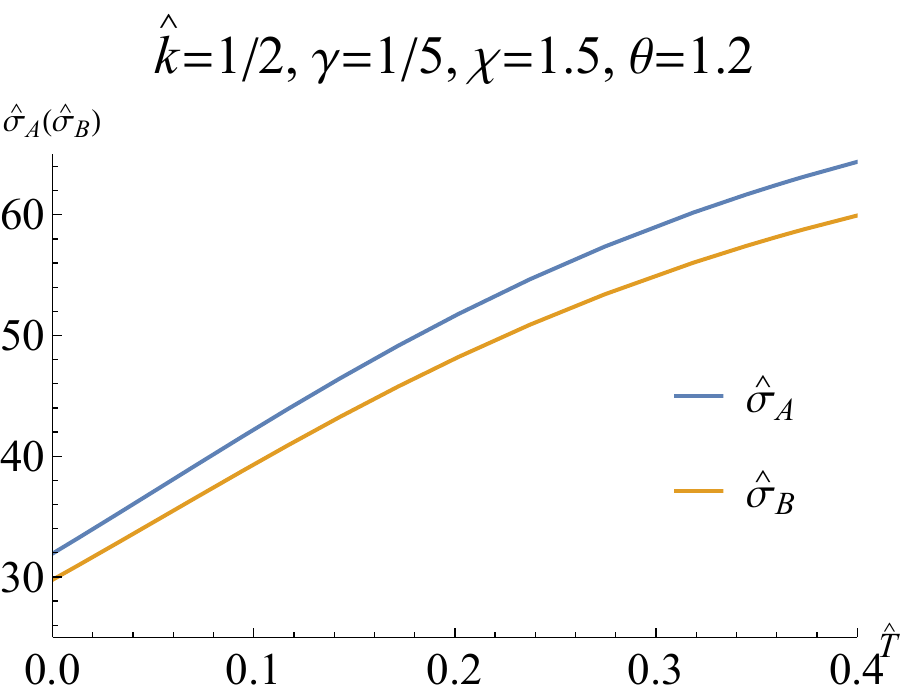}}
	\caption{\label{sigmaA-B1-theta} Electric conductivities $\hat{\sigma}_{A}$ and $\hat{\sigma}_{B}$ as a function of the temperature $\hat{T}$ . Here, we've specified $\hat{k}=1/2, \gamma=1/5, \chi=0.5,1.5, \theta=-0.2,0.5,1.2$.}
\end{figure}
Furthermore, it is discovered that the effects of doping $\chi$ and coupling $\theta$ on the spin-spin conductivity are similar to those of electric conductivity. However, it is fascinating to investigate the relative changes in electric and spin-spin conductivities, as shown in Fig.\ref{sigmaA-B1-theta}. It is easy to find that at a fixed $\chi$, as $\theta$ increases, the two conductivity curves progressively approach each other, and eventually coincide when $\theta=1$. When $\theta$ exceeds $1$, we observe that as $\theta$ increases, the two conductivity curves begin to separate from each other. When we change $\chi$, we detect comparable changes between the electric and spin-spin conductivities for the fixed coupling parameter $\theta$ (also see Fig.\ref{sigmaA-B1-theta}). 

\subsection{Thermal conductivities}
We're also interested in the properties of thermal transports. In addition to the thermal conductivity $\bar{\kappa}$ defined in Eq.\eqref{bar2}, we also introduce another thermal conductivity at zero electric current:
\begin{eqnarray}
	\label{kappaA-d}
	\kappa_{A}\equiv\bar{\kappa}-\frac{\alpha^2 T}{\sigma_{A}}\,,
\end{eqnarray}
which is more readily measurable than $\bar{\kappa}$. Then we want to express both thermal conductivities
in terms of the black hole entropy density $s$ and its charges $q_A$, $q_B$, which are the intrinsic quantities, as follows
\begin{eqnarray}
	&&
	\label{kappab-s}
	\bar{\kappa}=\frac{(s+2k^2\pi \gamma)^2 T}{M_{h}^{2}} \,, 
	\ \nonumber
	\\
	&&
	\label{kappaA-s}
	\kappa_{A}=\frac{(s+2k^2\pi \gamma)^2 T}{M_{h}^{2}+(q_A+\theta q_B)^{2}}\,.
\end{eqnarray}	
Here, the black hole entropy density $s$ can be calculated by the Wald formula as $s=4\pi r_h^2(1-\frac{\gamma}{2r_h^2}k^2)$ \cite{Baggioli:2017ojd}. Obviously, it relies on the Horndeski parameter $\gamma$.

Some comments on both thermal conductivities are presented as follows:
\begin{itemize}
	\item The thermal conductivities $\bar{\kappa}$ and $\kappa_{A}$ are affected not only by the intrinsic quantities, such as the black hole entropy density  $s$ and its charges $q_A$, $q_B$, but also by the model parameters $k$, $\gamma$ and $\theta$. When these model parameters tend to zero, the thermal conductivities reduce to those of Einstein-Maxwell theory, which are totally governed by the black hole's instrinsic quantities.
	\item $\bar{\kappa}$ is affected by the Horndeski parameter $\gamma$ but not by the coupling $\theta$, which describes the coupling between the two gauge fields. While $\kappa_{A}$ depends on both $\gamma$ and $\theta$. Recalling that in \cite{Wu:2018zdc}, the thermal conductivity at zero current is also affected by the non-linear Maxwell parameter, i.e., the Born-Infeld (BI) parameter, while the usual thermal conductivity $\bar{\kappa}$ is unaffected by this BI parameter.
	\item $\kappa_A$ is finite in the small momentum dissipation limit, i.e., $k \to 0$, whereas $\bar{\kappa}$ diverges. It indicates that, even in the limit of $k \to 0$,  $\kappa_A$ is also a well-defined quantity as compared to $\bar{\kappa}$.
\end{itemize}

\subsection{Lorentz ratios}

Fermi liquid has a noteworthy property \cite{ziman2001electrons}: the Lorenz ratio of thermal conductivity to electric conductivity remains constant at low temperatures. This property is dubbed as Wiedemann-Franz (WF) law. It may be computed directly for the Fermi liquid at low temperature as $L^{FL}\equiv\frac{\kappa}{\sigma T}=\frac{\pi^2}{3}$ in units with $k_B=e=1$. However, the WF law is fragile. This law has been found to be broken in the strongly interacting non-Fermi liquids \cite{Mahajan:2013cja,PhysRevLett:102156404}. Because of the inelastic scattering between charged and neutral degrees of freedom, it can be ascribed to heat and charge transport in different ways \cite{Mahajan:2013cja}. 
Furthermore, it has also been discovered that the WF law is also violated in the majority of holographic dual systems \cite{Kuang:2017rpx,Kim:2014bza,Donos:2014cya,Wu:2018zdc,Li:2022yad}. However, the mechanism behind them is still missing. We expect that by analyzing the properties of the Lorentz ratios in our current model, we may be able to give some insights to address this issue in the future.
Of this section, we will look more closely at the Lorentz ratio features in our current holographic model. We would like to mention that even for the holographic two-currents model without Horndeski coupling, the Lorentz ratio features are also absent. 

\begin{figure}[ht!]
	\centering
	\subfigure[]{
		\includegraphics[width=6.6cm]{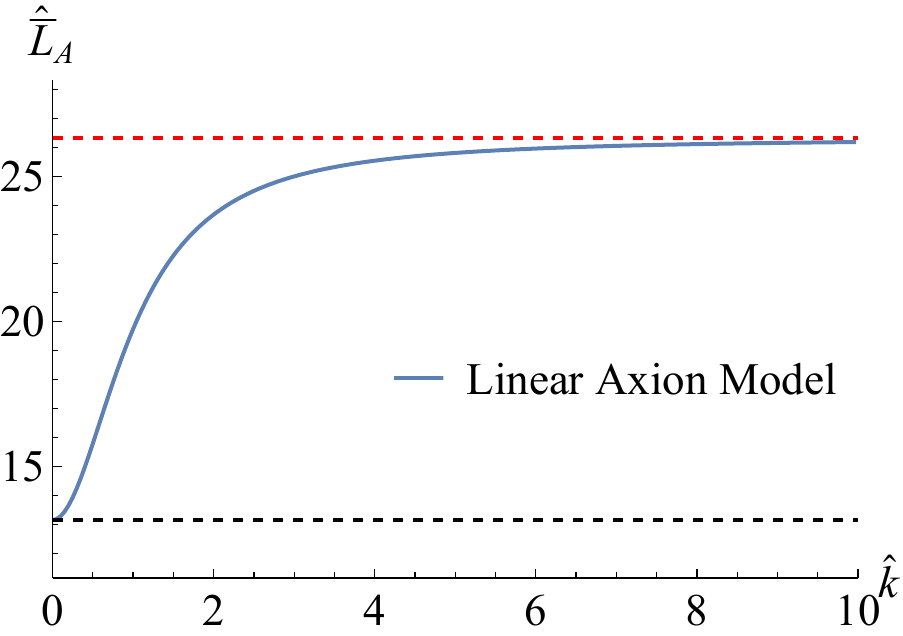}}
	\hspace{0.6in}
	\subfigure[]{
		\includegraphics[width=6.6cm]{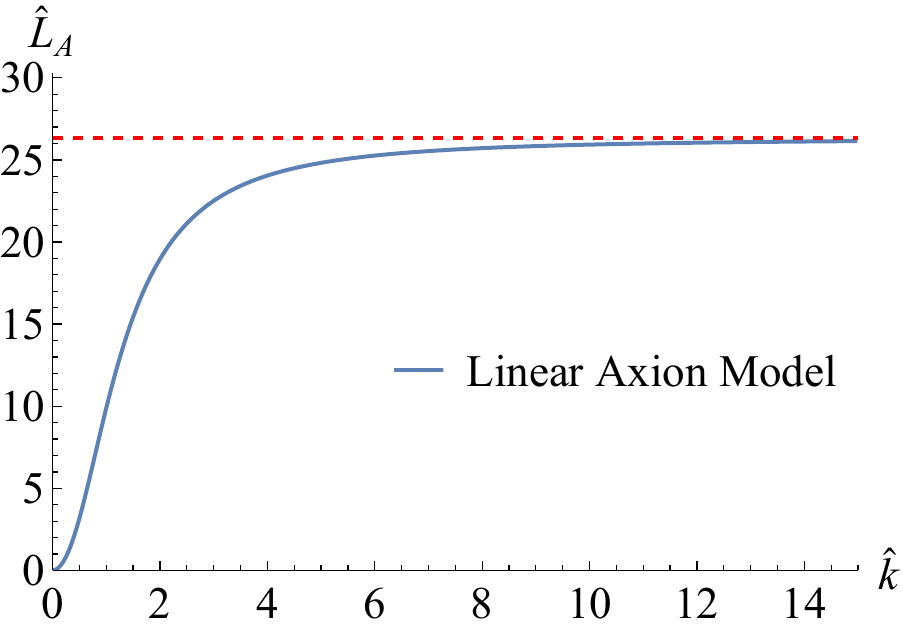}}
	\caption{\label{LA_RN} The Lorentz ratios $\hat{L}_A$ and $\hat{\bar{L}}_A$ as a function of $\hat{k}$ in the zero temperature limit for the typical linear axions model. The red dashed line indicates the upper bound given by the limit $\hat{k}\gg 1$, while the black dashed line represents the lower bound set by the limit $\hat{k} \ll 1$.}
\end{figure}

Before we go any further, let's go through the major aspects of the Lorentz ratios for the typical linear axions model. We are interested in the scenario of the zero temperature limit, where the Lorentz ratios can be written as \cite{Donos:2014cya,Kim:2014bza}
\begin{eqnarray}
&&
\label{LbarA-axion}
\hat{\bar{L}}_A\Big{|}_{\hat{T}\to 0}\equiv\frac{\hat{\bar{\kappa}}}{\hat{\sigma} \hat{T}}=\frac{16\pi^2}{\mu^2(\hat{k}^2+1)} \,, 
\
\\
&&
\label{LA-axion}
\hat{L}_A\Big{|}_{\hat{T}\to 0}\equiv\frac{\hat{\kappa}}{\hat{\sigma} \hat{T}}=\frac{16\pi^2\hat{k}^2}{\mu^2(\hat{k}^2+1)^2}\,.
\end{eqnarray}

To visualize this picture, we show the Lorentz ratios $\hat{L}_A$ and $\hat{\bar{L}}_A$ as a function of $\hat{k}$ in Fig.\ref{LA_RN}.
We see that the Lorentz ratios increases as the momentum dissipation increases. As a result, the WF law is broken as expected. However, it is discovered that there exist two bounds, the upper bound and the lower bound, which are established by two extremal limits, $\hat{k}\ll 1$ and $\hat{k}\gg 1$, respectively. Furthermore, we may work out these two bounds explicitly in both extreme limits:
\begin{eqnarray}
	&&
	\hat{\bar{L}}_A =  \left\{\begin{array}{cc}
		\frac{4 \pi^{2}}{3}\,,\,\,\,\,\,\,\,\,\,\hat{k}\ll 1 \\
		\frac{8 \pi^{2}}{3}\,,\,\,\,\,\,\,\,\,\,\hat{k}\gg 1
	\end{array}\right. \,, 
\
\\
&& 
	\hat{L}_A =  \left\{\begin{array}{cc}
	0\,,\,\,\,\,\,\,\,\,\,\,\,\,\, \hat{k}\ll 1\ \\
	\frac{8 \pi^{2}}{3}\,,\,\,\,\,\,\,\,\,\,\hat{k}\gg 1
\end{array}\right. .
\end{eqnarray}
We discover that, while the Lorentz ratios tend to constants in these extreme limits, the values differ from the Fermi-liquid case. It indicates that holographic systems are the ones with strongly interaction, akin to the non-Fermi liquid theory \cite{Mahajan:2013cja,PhysRevLett:102156404}.

\begin{figure}[ht!]
	\centering
	\subfigure[]{
		\includegraphics[width=6.5cm]{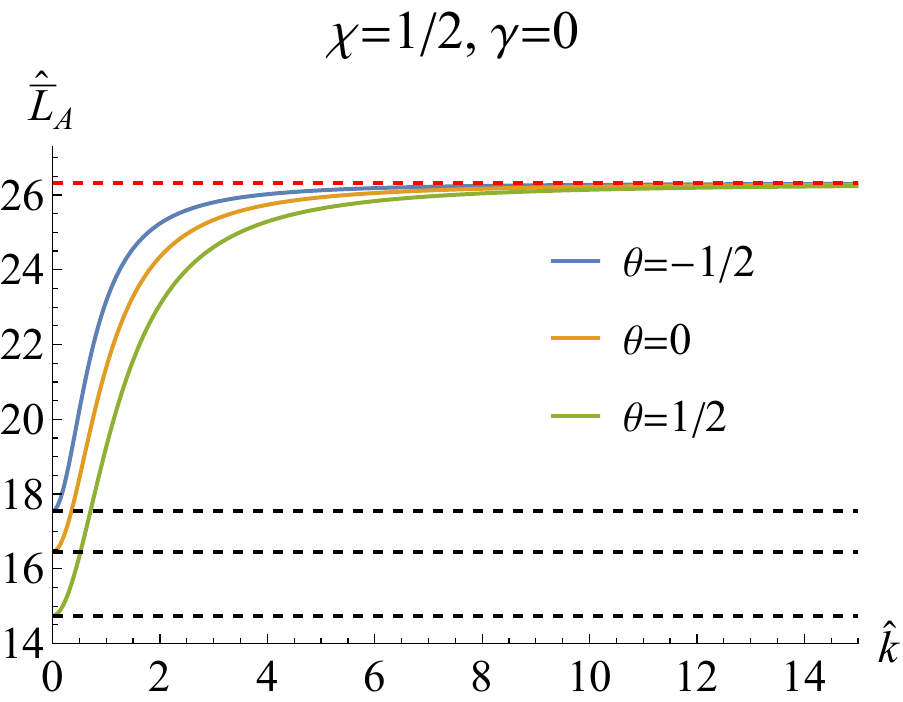}}
	\hspace{0.5in}
	\subfigure[]{
		\includegraphics[width=6.5cm]{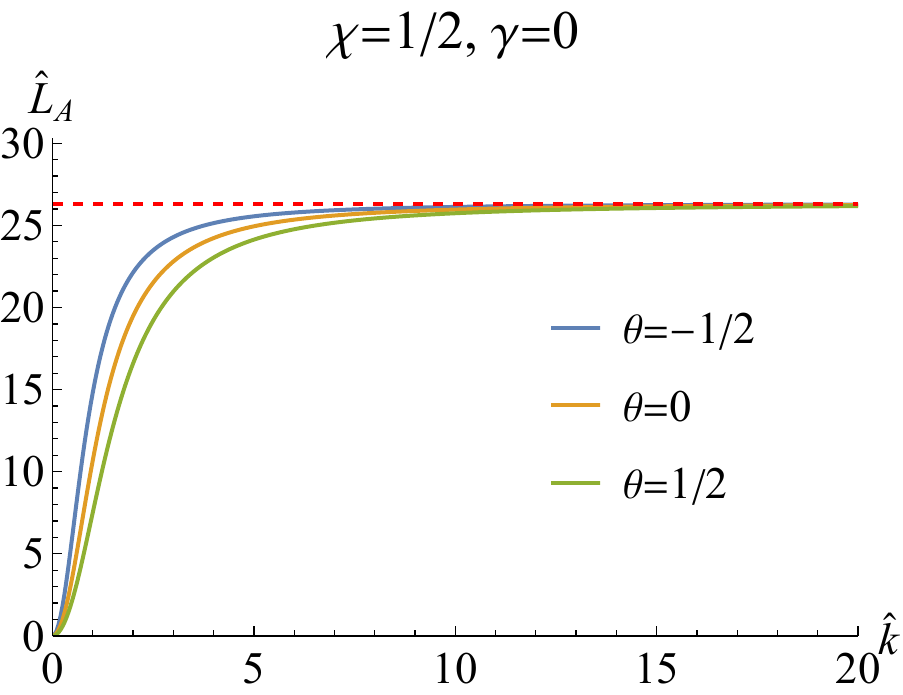}}
	\caption{\label{LA_gamma_0} The Lorentz ratios $\hat{L}_A$ and $\hat{\bar{L}}_A$ as a function of $\hat{k}$ for the holographic two-currents model. The red dashed line represents the upper bound established by the limit $\hat{k}\gg 1$, while the black dashed line represents the lower bound set by the limit $\hat{k} \ll 1$.}
\end{figure}
\begin{figure}[ht!]
	\centering
			\includegraphics[width=7cm]{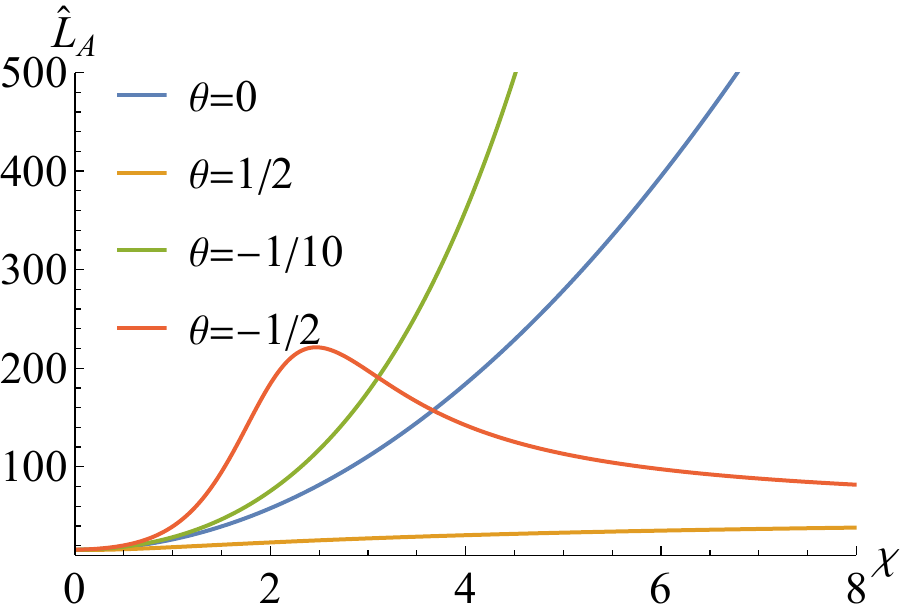}
	\caption{\label{LABARchi-fig} The Lorentz ratios $\hat{\bar{L}}_A$ as a function of $\chi$ for different $\theta$. Here we have set $\hat{k}=1/2$, $\gamma=0$.}
\end{figure}
Then we turn to the holographic two-currents model without Horndeski coupling, for which the Lorentz ratios $\hat{L}_A$ and $\hat{\bar{L}}_A$ in the zero temperature limit can be generalized to be\footnote{Here we discuss solely the Lorentz ratios for gauge field $A$, which we label as $\hat{L}_A$ and $\hat{\bar{L}}_A$.}
\begin{eqnarray}
	&&
	\label{LAb-2current}
	\hat{\bar{L}}_A\Big{|}_{\hat{T}\to 0}=\frac{16\pi^2}{\mu^2(\hat{k}^2+(1+\theta \chi)^2)} \,,  
	\ 
	\\ 
	&&
	\label{LA-2current}
	\hat{L}_A\Big{|}_{\hat{T}\to 0}=\frac{16\pi^2\hat{k}^2}{\mu^2(\hat{k}^2+(1+\theta\chi)^2)^2}\,.
\end{eqnarray}
We can observe that the system parameters $\theta$ and $\chi$ have an impact on the Lorentz ratios. Fig.\ref{LA_gamma_0} shows the Lorentz ratios as a function of $\hat{k}$ for various different coupling parameters $\theta$ and the fixed $\chi$.  
We have observed that the holographic two-currents model follows the same pattern as the typical axions model, where the Lorentz ratios increase with increasing strength of momentum dissipation.
However, the lower bound of $\hat{\bar{L}}_A$ varies with the system parameters $\theta$ and $\chi$ (left-plot in Fig.\ref{LA_gamma_0}). In particular, in the small momentum dissipation region, we have illustrated the variation of  $\hat{\bar{L}}_A$ as a function of $\chi$ for different $\theta$ in Fig.\ref{LABARchi-fig}. We confirm the observation in \cite{Seo:2016vks} that the existence of a new current can significantly violates the WF law. Additionally, we have observed that the kinetic mixing term either increases or decreases the heat transport relative to the charge transport depending on the coupling $\theta$. This finding provides us with the opportunity to finely adjust the coupling parameters and accurately model real-world condensed matter phenomena, such as graphene.

In addition, we compute the bounds in both extreme limits, $\hat{k}\ll 1$ and $\hat{k}\gg 1$, analytically, as follows:
\begin{eqnarray} 
	&&
	\label{LAb_2bound}
	\hat{\bar{L}}_A =  \left\{\begin{array}{cccc}
		\frac{4 \pi^{2}}{3} \frac{1+2\theta\chi+\chi^2}{(1+\theta\chi)^2}\,,\,\,\,\,\,\,\,\,\,\hat{k}\ll 1 \\
		\frac{8 \pi^{2}}{3}\,,\,\,\,\,\,\,\,\,\,\,\,\,\,\,\,\,\,\,\,\,\,\,\,\,\,\,\,\,\,\,\,\hat{k}\gg 1
	\end{array}\right. \,, \ \\
&&
\label{LA_2bound} 
	\hat{L}_A =  \left\{\begin{array}{cccc}
		0\,,\,\,\,\,\,\,\,\,\,\,\,\,\,\,\,\,\,\,\,\,\,\,\,\,\,\,\,\,\,\,\,\,\,\,\,\hat{k}\ll 1 \\
		\frac{8 \pi^{2}}{3}\,,\,\,\,\,\,\,\,\,\,\,\,\,\,\,\,\,\,\,\,\,\,\,\,\,\,\,\,\,\,\,\,\hat{k}\gg 1
	\end{array}.\right.
\end{eqnarray}
This analytical result backs with our observation from Fig.\ref{LA_gamma_0}. Aside from the lower bound of $\hat{\bar{L}}_A$, the other bounds are the same as for the typical axions model and are independent of the two-currents model parameters.

\begin{figure}[ht!]
	\centering
	\subfigure[]{
		\includegraphics[width=6.5cm]{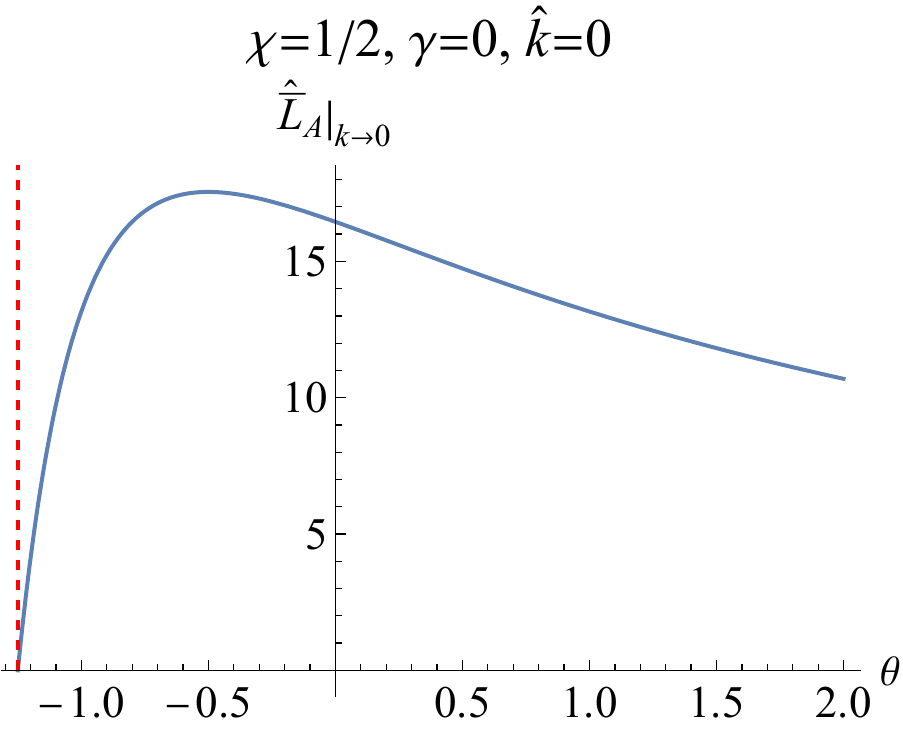}}
	\hspace{0.5in}
	\subfigure[]{
		\includegraphics[width=6.5cm]{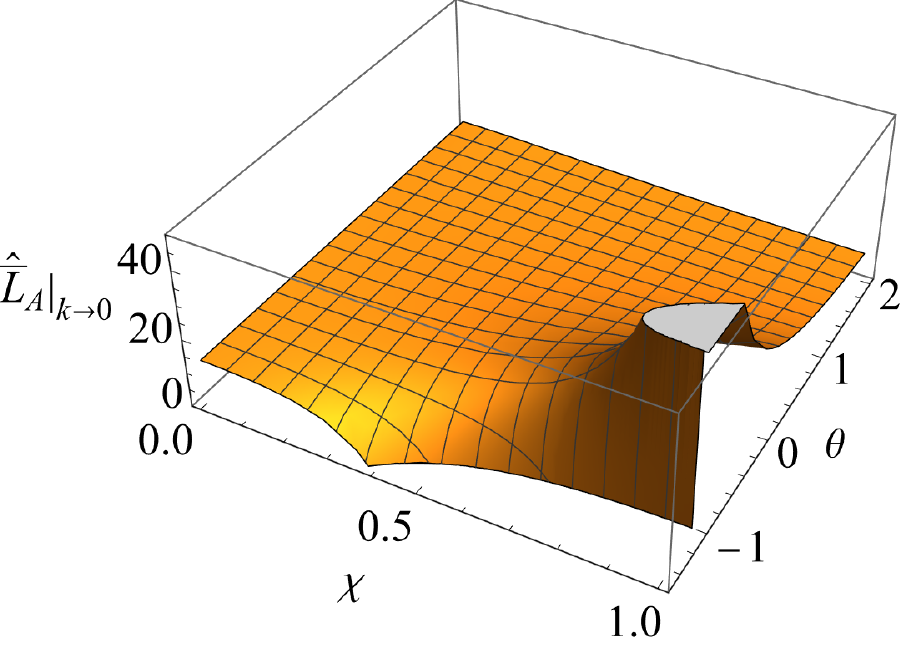}}
	\caption{\label{LAb_gamma_0_3d} Left plot: The lower bound as a function of $\theta$ for fixed $\chi$.
		Right plot: 3D plot of the lower bound as a function of $\chi$ and $\theta$.
}
\end{figure}
We are particularly interested in the lower bound of $\hat{\bar{L}}_A$. Left plot in Fig.\ref{LAb_gamma_0_3d} shows this bound as a function of $\theta$ for the fixed $\chi=1/2$. We find that the behavior of this bound is nonmonotonic with $\theta$. We especially notice that when $\theta$ is less than a certain value, the Lorentz ratio might be negative, which is usually forbade. 
We also present a 3D visualization of the lower bound as a function of $\chi$ and $\theta$. It is obvious that there are
certain locations where the Lorentz ratio is negative. Simultaneously, we see certain infinite peaks for some system parameters. To keep the Lorentz ratio positive and free of divergence, we can impose the following conditions: $1+2\theta \chi+\chi^2>0$ and $\theta\chi \neq -1$.

\begin{figure}[ht!]
	\centering
	\subfigure{
		\includegraphics[width=6.5cm]{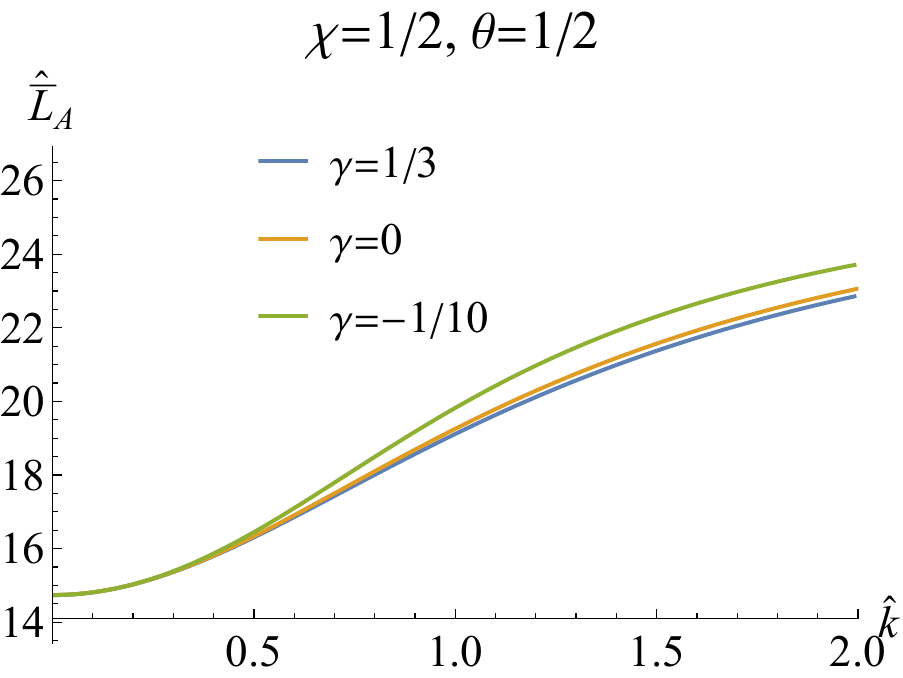}}
	\caption{\label{LA_gamnma} The Lorentz ratios  $\hat{\bar{L}}_A$ as a function of $\hat{k}$ for different Horndeski parameter $\gamma$.}
\end{figure}

We will now investigate the effect of the Horndeski coupling. Using the same approach, we may obtain the Lorentz ratio expressions in Horndeski theory as follows:
\begin{eqnarray}
	&&
	\label{barL}
\hat{\bar{L}}_{A}=\frac{64 \pi^{3} E_f^{2}}{
	\hat{k}^{2} \mu^{2}\left(2 \pi\left(\gamma\left(\hat{k}^{2} \mu^{2}-6\right)+2\right) E_f^{2}+\gamma \mu^{2}\left(4(\theta \chi+1)^{2}+\gamma \hat{k}^{2} \mu^{2}\left(2 \theta \chi+\chi^{2}+1\right)\right)\right)}\,,
\ \nonumber
\\
&&
\label{LA}
\hat{L}_{A}=\frac{64 \pi^{3} E_f^2\left(2 \pi\left(\gamma\left(\hat{k}^{2} \mu^{2}-6\right)+2\right) E_f^2+\gamma^{2} \hat{k}^{2} \mu^{4}\left(2 \theta \chi+\chi^{2}+1\right)\right)}{\left(2\pi \hat{k} \mu\left(\gamma\left(\hat{k}^{2} \mu^{2}-6\right)+2\right) E_f^2+\gamma \hat{k} \mu^{3}\left(4(\theta \chi+1)^{2}+\gamma \hat{k}^{2} \mu ^{2}\left(2 \theta \chi+\chi^{2}+1\right)\right)\right)^2}\,, \nonumber \\
\end{eqnarray}
where $E_f=efri\left(\frac{1}{2}\sqrt{\gamma}\hat{k}\mu\right)$. 
Both formulations are tedious, preventing intuitive insight, and we would want to expand them in the small $\gamma$ limit to the following forms:
\begin{eqnarray}
	\label{barL_E}
	\hat{\bar{L}}_{A}=&&\frac{16\pi^2\hat{k}^2}{\mu^2(\hat{k}^2+(1+\theta \chi)^2)^2}-\frac{4\pi^2\hat{k}^2\gamma}{3\mu^2(\hat{k}^2+(1+\theta\chi)^2)^3} \nonumber \\
	&&\left(6\hat{k}^4\mu^2-3(1+\theta\chi)^2(\mu^2(1+2\theta\chi+\chi^2)-12)-\hat{k}^2(36+\mu^2(7+\chi(2\theta(7+5\theta\chi)-3\chi)))\right)
	 \nonumber \\
	 &&
	+\mathcal{O}(\gamma^2)\,,
	\
	\\
	\label{LA_E}
	\hat{L}_{A}=&&\frac{16\pi^2\hat{k}^2}{\mu^2(\hat{k}^2+(1+\theta \chi)^2)}-\frac{4\hat{k}^2\pi^2\gamma(\mu^2(1+6\hat{k}^2+2\theta\chi+3\chi^2-2\theta^2\chi^2))}{3\mu^2(\hat{k}+(1+\theta\chi)^2)^2}+\mathcal{O}(\gamma^2)\,.
\end{eqnarray}
It is self-evident that the Horndeski coupling parameter $\gamma$ always appears in pairs with $\hat{k}$.
It means that the parameter $\gamma$ has no effect on the Lorentz ratio bounds. Furthermore, we depict the Lorentz ratios as a function $\hat{k}$ for various $\gamma$ (Fig.\ref{LA_gamnma}). It validates the finding that when $\gamma$ is small, it has no effect on the Lorentz ratio bounds.

\section{Conclusion and discussion}\label{conclusion}

We investigate the transport features of the holographic two-currents model in the Horndeski gravity framework in this research. The DC conductivities, including the electric and spin-spin conductivities associated with both gauge fields, the thermo-electric and thermo-spin conductivities, and the thermal and spin conductivities, are derived. Then, we primarily study the properties of the electric and spin-spin conductivities.
An interesting characteristic is that this holographic system exhibits metallic or insulating behaviors depending on whether the Horndeski parameter $\gamma$ is negative or positive, but is independent of other system parameters such as  momentum dissipation strength $\hat{k}$, doping $\chi$ and coupling $\theta$. 
Furthermore, we discover that doping $\chi$ and coupling $\theta$ have comparable effects on the spin-spin conductivity as they do on electric conductivity.

We also look at the thermal conductivities $\bar{\kappa}$ and $\kappa_{A}$ briefly. These thermal conductivities, as we know, may be determined by the intrinsic quantities, the black hole entropy density and its charges, in the typical axions model. However it is discovered that in the Horndeski framework that thermal conductivities are affected not only by intrinsic quantities but also by model parameters.

The Lorentz ratios' properties are then investigated. We pay special attention to the scenario of the zero temperature limit. In the holographic two-currents model without Horndeski coupling, the system parameters $\theta$ and $\chi$ both impact the Lorentz ratios and the WF law is broken.
By studying the case in the extremal limits, $\hat{k}\ll 1$ and $\hat{k}\gg 1$, we discover the upper and lower bounds of the Lorentz ratio $\hat{L}_A$, and the upper bound of $\hat{\bar{L}}_A$. These bounds are the same as the typical axions model. Of special interest is the lower bound of $\hat{\bar{L}}_A$, which relies on the doping parameter $\chi$ and the coupling parameter $\theta$. It is different from the case of the typical axions model. 
To keep the Lorentz ratio positive and free of divergence, the following requirements must be met: $1+2\theta \chi+\chi^2>0$ and $\theta\chi \neq -1$.
Furthermore, in the Horndeski framework, the coupling parameter $\gamma$ is always found in pairs with $\hat{k}$. It suggests that the Horndeski coupling parameter has no effect on the Lorentz ratio bounds.

Recalling that the real part of AC conductivity exhibits a dip at low frequency in the holographic two-currents model without momentum dissipation \cite{Zhang:2020znl}. In particular, a soft gap with power law decay emerges in the low frequency region. As a result, it is intriguing to investigate the AC conductivities in our current model further, and it is expected that some novel phenomena will emerge. 
Furthermore, it would be very worthwhile to study incoherent transports in holographic two-currents model with momentum dissipation, and further in the Horndeski gravity framework, in order to address the roles of doping and coupling $\theta$ in low-frequency transports. It is demonstrated that the high derivative term generally violates the diffusivity bounds, see for example \cite{Li:2017nxh,Mokhtari:2017vyz}. It will be fascinating to see if the diffusivity bounds hold in our current model. We will return to these topics in the near future.
	
\acknowledgments

We are very grateful to Zhenhua Zhou for helpful discussions and suggestions.
This work is supported by the Natural Science Foundation of China under Grants Nos. 11775036 and 12275079, and the Postgraduate Research \& Practice Innovation Program of Jiangsu Province under Grant No. KYCX20\_2973 and KYCX21\_3192, and the Postgraduate
Scientific Research Innovation Project of Hunan Province under Grant No. CX20220509. J.-P.W. is also supported by Top Talent Support Program from Yangzhou University.	
	
\appendix
\section{Equations of motion}\label{appendix}
We will derive the EOMs for this model in this appendix. Applying the variational approach to the action \eqref{Action}, the EOMs are derived as:
\begin{eqnarray}\label{covariant form}
&&
\label{EE}
G_{\mu\nu}+\Lambda g_{\mu\nu}-\frac{1}{2}\mathcal{T}^{(A)}_{\mu\nu}-\frac{1}{2}\mathcal{T}^{(B)}_{\mu\nu}-\theta\mathcal{T}^{(AB)}_{\mu\nu}
-\frac{1}{2}\mathcal{T}^{(\phi)}_{\mu\nu}-\sum_{I=1}^2\frac{\gamma}{2}\mathcal{T}^{G}_{_{\mu\nu}}=0\,,
\
\\
&&
\nabla_{\mu}[(g^{\mu\nu}-\gamma G^{\mu\nu})\partial_{\nu}\phi^{I}]=0\,,
\label{eom-axion}
\\
&&
\nabla_{\mu}(F^{\mu\nu}+\theta Y^{\mu\nu})=\nabla_{\mu}(Y^{\mu\nu}+\theta F^{\mu\nu})=0\,,
\label{ME}
\end{eqnarray}
where energy-momentum tensors $\mathcal{T}^{(\phi)}_{\mu\nu}, \mathcal{T}^{(A)}_{\mu\nu}, \mathcal{T}^{(B)}_{\mu\nu}, \mathcal{T}^{(AB)}_{\mu\nu}$ and  $\mathcal{T}^{(G)}_{\mu\nu}$ in the Einstein equation \eqref{EE} are defined as
\begin{eqnarray}\label{energy-momentum tensor}
\label{eom}
\mathcal{T}^{(A)}_{\mu\nu}&=&F_{\mu\rho}F_{\nu}^{\ \rho}-\frac{1}{4}g_{\mu\nu}F^2\,,\nonumber\\
\mathcal{T}^{(B)}_{\mu\nu}&=&Y_{\mu\rho}Y_{\nu}^{\ \rho}-\frac{1}{4}g_{\mu\nu}Y^2\,,\nonumber\\
\mathcal{T}^{(AB)}_{\mu\nu}&=&F_{(\mu|\rho|}Y_{\nu)}^{\ \rho}-\frac{1}{4}g_{\mu\nu}F_{\alpha\beta}Y^{\alpha\beta}\,,\nonumber\\
\mathcal{T}^{(\phi)}_{\mu\nu}&=&\sum_{I=1}^2(\partial_{\mu}\phi^{I}\partial_{\nu}\phi^{I}-\frac{1}{2}g_{\mu\nu}(\partial\phi^{I})^{2})\,,\nonumber\\
\mathcal{T}^{(G)}_{\mu\nu}&=&\frac{1}{2}\partial_{\mu}\phi^{I}\partial_{\nu}\phi^{I}R- 2\partial_{\rho}\phi^{I}\partial_{ ( \mu} \phi^{I}R_{\nu)}\displaystyle^{\rho}
-\partial_{\rho}\phi^{I}\partial_{\sigma}\phi^{I}R_{\mu}\displaystyle^{\rho}\displaystyle_{\nu}\displaystyle^{\sigma}\nonumber \\
&&-(\nabla_{\mu}\nabla^{\rho}\phi^{I})(\nabla_{\nu}\nabla_{\rho}\phi^{I})+ (\nabla_{\mu}\nabla_{\nu}\phi^{I})\Box\phi^{I}+\frac{1}{2}G_{\mu\nu}(\partial\phi^{I})^{2}\nonumber\\
&&-g_{\mu\nu}[-\frac{1}{2}(\nabla^{\rho}\nabla^{\sigma}\phi^{I})(\nabla_{\rho}\nabla_{\sigma}\phi^{I})+
\frac{1}{2}(\Box\phi^{I})^{2}-\partial_{\rho}\phi^{I}\partial_{\sigma}\phi^{I}R^{\rho\sigma}] \,.
\end{eqnarray}
Substituting the ansatz \eqref{ansatz} into the above EOMs, we find that the equation of axions field \eqref{eom-axion} is automatically satisfied. The Einstein and Maxwell equations can be written down explicitly
\begin{eqnarray}\label{EOMs}
&&
f'+\left(\frac{1}{r}-\frac{k^2 \gamma}{2r^3}\right)f+\frac{rf\mathcal{F}(A_t,B_t)}{4h}+\frac{k^2-6r^2}{2r}=0\,, \  \\
&&
h'+\frac{(k^2 \gamma f +2r^2 f +k^2 r^2-6r^4)}{2r^3f}h+\frac{r\mathcal{F}(A_t,B_t)}{4}=0\,, \  \\
&&
h''+\frac{r^2+k^2 \gamma}{r^3}h'-\frac{(12r^4+2k^2\gamma f-(2r^3+k^2r\gamma)f')}{2r^4f} h-\frac{1}{2}\mathcal{F}(A_t,B_t)=0\,,\ \\
&&
A_t''+\frac{1}{2}\left(\frac{4}{r}+\frac{f'}{f}-\frac{h'}{h}\right)A_t'=0\,,\  \\
&&
B_t''+\frac{1}{2}\left(\frac{4}{r}+\frac{f'}{f}-\frac{h'}{h}\right)B_t'=0\,,
\end{eqnarray}	
where $\mathcal{F}(A_t,B_t)=A_t'^2+2\theta A_t'B_t'+B_t'^2$ and the functions $h\,,\ f\,, \ A_t$ and $B_t$ only depend on the radial direction $r$.

\section{Derivation of DC conductivity}\label{appendix-B}

In this appendix, we demonstrate the procedure for calculating DC conductivity using the method proposed in \cite{Donos:2014cya}. The key point of this method is to build a radially-conserved current that connects the horizon and the boundary. Thus, the DC conductivity of the dual boundary system can be read off by the horizon datas directly. 

The consistent perturbations around the background are given by
\begin{eqnarray}\label{ peturbations form}
    &&
    g_{tx}= H(r)t+r^2 \delta h_{tx}(r)\,,~~~ g_{rx}=r^2\delta h_{rx}(r)\,,~~~ \phi_{x}=kx+ \delta\chi_{x}(r)\,,
    \
    \\
    &&
    A_{x}=E_{Ap} t+\delta a_x(r)\,,~~~ B_{x}=E_{Bp} t+\delta b_x(r)\,,\ \\
    &&
    H(r)=-\zeta h(r)\,,~~~ E_{Ap}=-E_{Ax}+\zeta A_t(r)\,,~~~ E_{Bp}=-E_{Bx}+\zeta B_t(r)\,,
\end{eqnarray}	
where $E_{Ax}$ and $E_{Bx}$ are external electric fields and $\zeta=-\nabla_{x} T/T$ is the temperature gradient. From the Maxwell and Einstein equations, one can construct radially-conserved currents in the bulk, which have the following forms:
\begin{eqnarray}\label{ peturbations form1}
    J_A^x&=&-\sqrt{-g}(F^{rx}+\theta Y^{rx})\,,~~~J_B^x =-\sqrt{-g}(Y^{rx}+\theta F^{rx})\,,
    \label{ current form}
    \\	
    Q^x&=&2\sqrt{-g}\nabla^rk^x-A_tJ_A^x-B_tJ_B^x\,,
    \label{ heat form}
\end{eqnarray}	
where $k^x$ is the killing vector ($k^x=\partial_t$). Furthermore, the electric and heat currents can be explicitly given by
\begin{eqnarray}\label{ peturbations form}
    J_A^x&=&-\sqrt{\frac{f}{h}}((H t+r^{2}\delta h_{tx})(A'_{t}+\theta B'_{t})+h((E_{Ap}'+\theta E_{Bp}')t+\delta a_x'+\theta \delta b_x'))\,,\\
    J_B^x&=&-\sqrt{\frac{f}{h}}((Ht+r^{2}\delta h_{tx})(B'_{t}+\theta A'_{t})+h((E_{Bp}'+\theta E_{Ap}')t+\delta b_x'+\theta \delta a_x'))\,,\\
    Q^x&=&h^{\frac{3}{2}}f^{\frac{1}{2}}\Big(\frac{g_{tx}}{h}\Big)'-A_tJ_A^x-B_tJ_B^x\,,
\end{eqnarray}	
where $A_{t}'=-\sqrt{\frac{h}{f}}\frac{q_{A}}{r^2}$ and $B_{t}'=-\sqrt{\frac{h}{f}}\frac{q_{B}}{r^2}$ have been taken into account. One can easily show that $\partial_r J_A^x=\partial_r J_B^x=\partial_r Q^x=0$. The regular boundary condition near the horizon requires that
\begin{eqnarray}\label{ infalling}
    \delta a_x'=\frac{E_{Ap}}{\sqrt{fh}}\,,~~~\delta b_x'=\frac{E_{Bp}}{\sqrt{fh}}\,,~~~\delta h_{tx}=\delta h_{rx}\sqrt{fh},
\end{eqnarray}	
along with the following constraint relation derived from the linearized Einstein equation
\begin{eqnarray}\label{ hrx }
    \delta h_{rx}=\frac{2(r(q_A+\theta q_B)E_{Ap} \sqrt{h}+r(q_B+\theta q_A)E_{Bp} \sqrt{h}+\sqrt{f}(2r^2 H+k^2\gamma H-r^3H'))}{r\sqrt{f}((q_A^2+q_B^2+2q_A q_B\theta -12r^4+4(r^2+k^2\gamma )f)h+2r(2r^2+k^2\gamma )fh')}.
\end{eqnarray}

From the generalized Ohm's law (\ref{dceq}) and the above relations, the DC conductivities are
\begin{align}\label{DC1-app-1}
    \sigma_{A}&=\frac{\partial J_{A}^x(r_{h})}{\partial E_{A_x}}=1+\frac{(q_A+\theta q_B)^2}{M_{h}^{2}} \,,	\\
    \sigma_{B}&=\frac{\partial J_{B}^x(r_{h})}{\partial E_{B_x}}=1+\frac{(q_B+\theta q_A)^2}{M_{h}^{2}} \,, \\
    \alpha &=\frac{1}{T}\frac{\partial J_{A}^x(r_{h})}{\partial \zeta }= \frac{4\pi (q_A+\theta q_B)}{M_{h}^{2} }\,, \\
    \beta &=\frac{1}{T}\frac{\partial J_{B}^x(r_{h})}{\partial \zeta}= \frac{4\pi (q_B+\theta q_A)}{M_{h}^{2} }\,, \\
    \eta&=\frac{\partial J_{A}^x(r_{h})}{\partial E_{B_x}}=\frac{\partial J_{B}^x(r_{h})}{\partial E_{A_x}}=\theta+\frac{(q_A+\theta q_B)(q_B+\theta q_A)}{M_{h}^{2}}\,,\\
    \label{bar2}
    \bar{\kappa}&=\frac{\partial Q(r_{h})}{\partial \zeta}=\frac{16\pi^{2}T}{M_{h}^{2}}\,,
\end{align}	
where $M_h^2$ is the effective graviton mass at the horizon
\begin{eqnarray}
    \label{Mh1}
    M_{h}^{2}=k^{2}(1-4e^{-\frac{k^{2}\gamma}{4}}\pi T \gamma)\,.
\end{eqnarray}	
One can also express the DC conductivities by dimensionless quantities denoted by the hat symbols
\begin{eqnarray}\label{DC1-app}
    \hat{\sigma}_{A}&=&1+\frac{\hat{k}^{2}\mu^{4}\gamma(1+\theta\chi)^{2}}{\hat{M_{h}^{2}}\pi erfi(\frac{\hat{k}\mu\sqrt{\gamma}}{2})^{2}} \,,	\\
    \hat{\sigma}_{B}&=&1+\frac{\hat{k}^{2}\mu^{4}\gamma(\theta+\chi)^{2}}{\hat{M_{h}^{2}}\pi erfi(\frac{\hat{k}\mu\sqrt{\gamma}}{2})^{2}} \,, \\
    \hat{\alpha} &=& \frac{4\hat{k}\mu^2\sqrt{\pi}\sqrt{\gamma}(1+\theta\chi)}{\hat{M_{h}^{2}} erfi(\frac{\hat{k}\mu\sqrt{\gamma}}{2})}\,, \\
    \hat{\beta} &=& \frac{4\hat{k}\mu^2\sqrt{\pi}\sqrt{\gamma}(\theta+\chi)}{\hat{M_{h}^{2}} erfi(\frac{\hat{k}\mu\sqrt{\gamma}}{2})}\,, \\
    \hat{\eta}&=&\theta+\frac{\hat{k}^{2}\mu^{4}\gamma(\theta+\chi)(1+\theta\chi)}{\hat{M_{h}^{2}}\pi erfi(\frac{\hat{k}\mu\sqrt{\gamma}}{2})^{2}}\,,\\
    \hat{\bar{\kappa}}&=&\frac{16\pi^{2}\hat{T}}{\hat{M_{h}^{2}}}\,,
\end{eqnarray}	
where 
\begin{eqnarray}
    \label{Mh}
    \hat{M_{h}^{2}}=\hat{k}^{2}\mu^{2}(1-4e^{-\frac{\hat{k}^{2}\mu^{2}\gamma}{4}}\pi \hat{T} \gamma\mu)\,.
\end{eqnarray}

\bibliographystyle{style1}
\bibliography{Ref}
\end{document}